\documentclass[final,1p,times,authoryear]{elsarticle}
\pdfoutput=1

\usepackage{fullpage}
\usepackage{amssymb}
\usepackage{amsmath}
\usepackage{lipsum}
\usepackage{nomencl}
\usepackage{pstricks}
\usepackage{tablefootnote}
\usepackage[colorlinks=true,linkcolor=blue,citecolor=blue,allcolors=blue]{hyperref}
\usepackage{siunitx}
\usepackage{cellspace}
\usepackage{amsmath, array, makecell}
\usepackage{multibib}
\usepackage[]{graphicx}
\usepackage{nomencl}
\makenomenclature

\renewcommand{\nomname}{List of Symbols}

\usepackage{lipsum}

\usepackage{units}
\usepackage[utf8]{inputenc}
\usepackage{multirow}
\usepackage{dsfont}

\usepackage{lineno}

\usepackage{setspace}
\makenomenclature

\journal{Journal of Computational Physics}

\begin{document}

\begin{frontmatter}



\title{Multiphase Flow Modelling in Multiscale Porous Media: An Open-Sourced Micro-Continuum Approach}

\author[PRINCETONCHEM]{Francisco J. Carrillo}
\author[PRINCETONCIVIL,PEI]{Ian C. Bourg}
\author[CNRS,BRGM]{Cyprien Soulaine}

\address[PRINCETONCHEM]{Department of Chemical and Biological Engineering, Princeton University, Princeton, NJ, USA}
\address[PRINCETONCIVIL]{Department of Civil and Environmental Engineering, Princeton University, Princeton, NJ, USA}
\address[PEI]{Princeton Environmental Institute, Princeton University, Princeton, NJ, USA}
\address[CNRS]{Earth Sciences Institute of Orl\'eans, Universit\'e d'Orl\'eans, CNRS, BRGM, Orl\'eans, France}
\address[BRGM]{French Geological Survey, BRGM, Orl\'eans, France}

\begin{abstract}
An open-sourced multiphase Darcy-Brinkman approach is proposed to simulate two-phase flow in hybrid systems containing both solid-free regions and porous matrices. This micro-continuum model is rooted in elementary physics and volume averaging principles, where a unique set of partial differential equations is used to represent flow in both regions and scales. The crux of the proposed model is that it tends asymptotically towards the Navier-Stokes volume-of-fluid approach in solid-free regions and towards the multiphase Darcy equations in porous regions. Unlike existing multiscale multiphase solvers, it can match analytical predictions of capillary, relative permeability, and gravitational effects at both the pore and Darcy scales. Through its open-source implementation, \textit{hybridPorousInterFoam}, the proposed approach marks the extension of computational fluid dynamics (CFD) simulation packages into porous multiscale, multiphase systems. The versatility of the solver is illustrated using applications to two-phase flow in a fractured porous matrix and wave interaction with a porous coastal barrier.
\end{abstract}


\end{frontmatter}


\section{Introduction}
\label{Sect:Intro}


Virtually all aspects of subsurface engineering for energy and environmental applications require in-depth understanding of multiphase flow within heterogeneous porous media. Examples include enhanced hydrocarbon recovery, geologic carbon sequestration, nuclear waste storage, geothermal energy production, seasonal storage of natural gas in geologic formations, and gas hydrate formation in sediments \citep{Lake2014,Li2015,Rocco2017,Yin2018}. In addition, multiphase fluid dynamics in heterogeneous porous media play key roles in the natural fluxes of water and carbon in soils and sediments \citep{Hassanizadeh2002,Or2013,Maxwell2014,Scandella2017} as well as in a variety of engineering processes \citep{Baber2016,Jabbari2016}. One largely unresolved challenge in the field is the inability to predict and characterize multiphase flow physics within inherently multiscale structures, particularly in systems that contain both porous and solid-free domains \citep{Helmig2013}. Although this challenge is widely recognized, there is increased urgency in addressing it because of the need to sequester billions of tons of CO$_2$ and to efficiently extract hydrocarbons without causing extensive environmental damage. 

Whereas single-phase flow in porous media is relatively well understood from atomistic to continuum scales, the dynamics of systems containing multiple phases remain challenging to describe at all scales \citep{Gray2015,Li2018}. Multiphase flow involves strong feedback between inertial, viscous, capillary, and interfacial forces \citep{Meakin2009,Datta2014}. This coupling is intrinsically multiscale, as inertial and viscous forces dominate at large pores or fractures while capillary forces and interfacial energetics dominate within smaller porous micro-structures. The complex linkage between microscopic geometric heterogeneities and macroscopic processes makes it necessary to consider scale-dependent processes across porous media in order to create truly predictive models: from the scale of microscopic interfaces ($\sim\mu$m), to the scale of pore-networks and lab columns ($\sim$cm), all the way up to the field scale ($\sim$km).

There are almost as many definitions of ``multiscale" as authors that invoke this concept. Nevertheless, multiscale modelling can be sorted in three main categories \citep{Helmig2010}: \emph{(i)} the multiscale homogenization strategy, \emph{(ii)} multiscale algorithm approach, and \emph{(iii)} the multiphysics approach. The first of these, the multiscale homogenization strategy, aims at deriving large scale models rooted in elementary physical principles by using homogenization techniques including volume averaging, mixture theory, and asymptotic expansions \citep{Whitaker1999,Standnes2017,Battiato2019,Starnoni2020}. A prime example is the seminal work of \cite{Whitaker1986}, which demonstrates that Darcy’s law arises from the integration of Stokes equation over a porous Representative Elementary Volume (REV). These upscaling techniques usually uncouple each scale's relevant physics through the scale-separation hypothesis. This way, effective coefficients in large-scale models can be used to describe fine-scale phenomena and geometric features. Such parameters are commonly estimated by using complementary fine-scale simulations on REVs or sub-grid models. The second strategy, the multiscale algorithm approach, solves flow physics on interconnected grids with different degrees of refinement. This way, each grid’s refinement level can be tuned to fit its respective scale of interest. A portion of these algorithms primarily focus on fine-scale solutions, and thus, use multi-scale finite volume/element solvers to speed up convergence in fine grids \citep{Jenny2003,Jenny2006,Efendiev2007}. Conversely, alternative algorithms focus on large-scale behaviors and only solve for small-scale behavior when needed \citep{Tomin2013,Tomin2016}. The third strategy, the multiphysics approach, uses domain decomposition to solve different physics within each scale’s sub-domain. In this method, sub-domains have their own independent set of governing equations and only interact with each other through the implementation of appropriate boundary conditions \citep{Sun2012a,Baber2016}. A popular implementation of this approach uses the \cite{Beavers1967} conditions to couple a porous domain governed by Darcy’s Law with a domain governed by the Navier-Stokes equations.

\begin{figure}
\begin{center}
\includegraphics[width=0.97\textwidth]{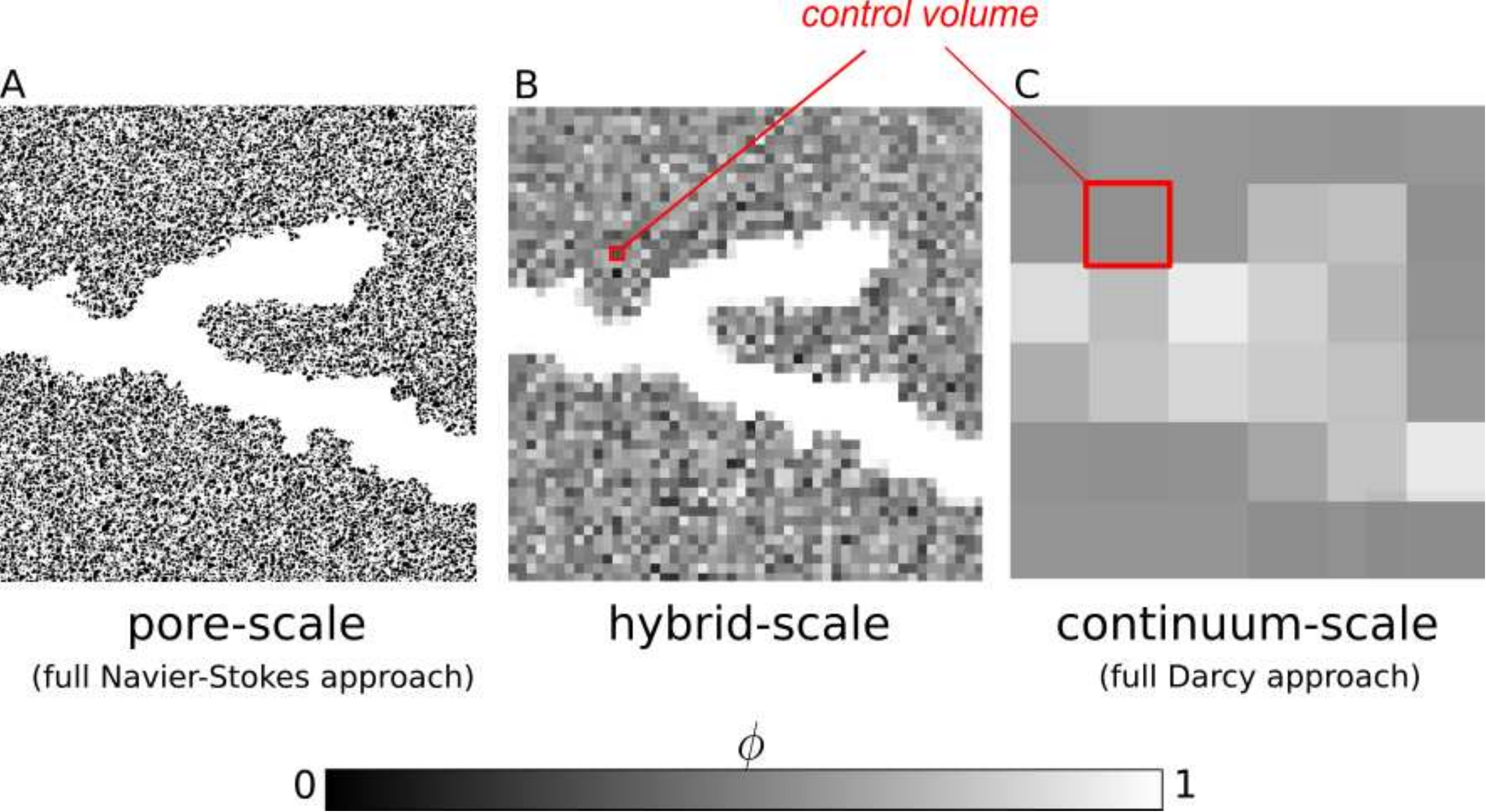}
\caption{Schematic representations of a porous medium with two characteristic pore sizes depending on the scale of resolution: (a) full pore scale (Navier-Stokes), (b) intermediate or hybrid scale, and (c) full continuum scale (Darcy). Our objective is to derive a framework that can describe multiphase flow at all three scales described in the figure based on a single set of equations resolved throughout the entire system.   \label{fig:Figure1}}
\end{center} 
\end{figure}

Here, we will implement concepts from all three strategies to propose an alternative solution to the multiscale challenge. To do so, we will rely on the micro-continuum approach \citep{Soulaine2016a}, whereby a single equation is used to handle flow and transport in systems where a large scale solid-free domain coexists with a small-scale porous domain (Figure \ref{fig:Figure1}). In the case of single-phase flow and transport, this approach generally relies on the well-known Darcy-Brinkman (DB) equation --also referred to as Darcy-Brinkman-Stokes (DBS) equation-- \citep{Brinkman1947} that arises from volume averaging the Stokes (or Navier-Stokes) equations in a control volume that contains both fluids and solids \citep{Vafai1981,Hsu1990,Bousquet-Melou2002,Goyeau2003}. It consists in a Stokes-like momentum equation that is weighted by porosity and contains an additional drag force term that describes the mutual friction between the fluids and solids within said control volume. Unlike standard continuum scale equations for flow and transport in porous media such as Darcy's law, the DB equation remains valid in solid-free regions (see Figure~\ref{fig:Figure1}A) where the drag force term vanishes and the DB equation turns into the Stokes (or Navier-Stokes) equation. In porous regions (see Figure~\ref{fig:Figure1}C), in contrast, viscous dissipation effects are negligible compared with the drag force exerted onto the pore walls and the DB momentum equation tends asymptotically towards Darcy's law \citep{Tam1969,Whitaker1986,Auriault2009}. Therefore, the micro-continuum DB equation has the ability to simultaneously solve flow problems through porous regions and solid-free regions \citep{Neale1974}, paving the path to hybrid scale modeling (see Figure~\ref{fig:Figure1}B). In the case of single phase flow, it is known to be analogous (in fact, formally equivalent) to the previously mentioned and well-established Beavers-Joseph boundary conditions \citep{Beavers1967,Neale1974}.

The ability of the DB equation to handle two scales simultaneously has been used to solve fluid flow in three-dimensional images of rock samples that contain unresolved sub-voxel porosity \citep{Knackstedt2006,Apourvari2014, Scheibe2015,Soulaine2016,Guo2018,Kang2019,Singh2019}. It also has been used to simulate dissolution wormholing during acid stimulation in cores by updating the weighting porosity field through geochemical reactions \citep{Liu1997,Golfier2002,Soulaine2016a,Tomin2018}. Moreover, it has been shown  that whenever low-porosity low-permeability porous regions are present, the velocity within these regions drops to near zero, such that the micro-continuum DB framework can be used as a penalized approach to map a solid phase onto a Cartesian grid with a no-slip boundary at the solid surface \citep{Angot1999,Khadra2000,Soulaine2016a}. This approach tends to a full Navier-Stokes representation of the flow physics at the pore scale and, hence, can be used to move fluid-solid interfaces efficiently in a Cartesian grid without a re-meshing strategy. For example, \cite{Soulaine2017} used a micro-continuum framework to predict the dissolution kinetics of a calcite crystal and successfully benchmarked their model against state-of-the-art pore scale dissolution solvers with evolving fluid-solid interfaces \citep{Molins2019}. Another example, presented in \cite{Carrillo2019}, leveraged this framework to create a Darcy-Brinkman-Biot approach capable of predicting the coupled hydrology and mechanics of soft porous media such as clays and elastic membranes.

The micro-continuum framework outlined above was limited, until recently, to single-phase flow \citep{Soulaine2016a}. \cite{Horgue2014} and later \cite{Soulaine2018} proposed the first two-phase micro-continuum model by combining a two-phase variant of the DB equation with the volume of fluid (VOF) approach \citep{Hirt1981} to two-phase flow in solid-free regions. This formulation enabled multiphase flow in solid-free regions with imposed wettability conditions at the solid surface while describing microporous regions as impervious, fully-saturated porous domains \citep{Soulaine2018}. \citet{Soulaine2019} later refined their formulation to enable two-phase flow at both the pore and continuum scales, but with simplified flow physics in the porous domain; in particular, the model could not describe the interplay of gravity and capillarity effects within the microporous matrix.

In this paper, we expand upon \citet{Soulaine2019} to propose a fully realized multiscale solver for two-phase flow in porous media rooted in elementary physical principles and rigorously derived using the method of the volume averaging \citep{Whitaker1999}. We show that there exists a single set of partial differential equations that can be applied in pore, continuum, or hybrid scale representation of multiphase flow in porous media. Particular attention is paid to the rigorous derivation of gravity and capillary effects in the porous domain. The resulting two-phase micro-continuum framework is verified using a series of test cases where reference solutions exist. We show that the multiscale solver converges to the standard Darcy scale solutions (Buckley-Leverett, capillary-gravity equilibrium, drainage in a heterogeneous reservoir) when used at the continuum scale and to the two-phase Navier-Stokes solutions (droplet on a flat surface, capillary rise, drainage with film deposition, two-phase flow in a complex porous structure) when used at the pore scale. The fully implemented numerical model, along with the aforementioned verification and tutorial cases, is provided as an open-source solver (\textit{hybridPorousInterFoam}) accompanying the present article.
 
The paper is organized as follows. In Section \ref{sec:Mathematical_model}, the multi-scale governing equations are rigorously derived using the method of volume averaging. Multi-scale parameters are then defined by asymptotic matching to the two-phase Navier-Stokes and Darcy equations. In Section \ref{sec:numerical_implementation}, we describe the numerical algorithm used to solve the problem (governing equations, constitutive relations, and boundary conditions) and present its numerical implementation as an open-source simulation platform. In Section \ref{sec:verification}, we present the model verification at the pore and continuum scales. In Section \ref{sec:applications}, we illustrate the versatility of the proposed framework by describing two hybrid scale applications: wave propagation in a coastal barrier and two-phase flow in a fractured porous matrix. We close with a summary and conclusions.

\section{Mathematical model \label{sec:Mathematical_model}}

In this section, we derive the micro-continuum equations for two-phase flow. First, we consider the conservation laws for multi-phase systems in the continuous physical space. Then, the micro-continuum equations are formed by volume averaging the continuous equations over each volume of an Eulerian grid. Finally, information below the size of the grid cell (fluid-fluid interface location and micro-structure geometry) is modelled with closure of the multiscale parameters.

\subsection{Governing equation in the continuous physical space}

\begin{figure}[h!]
\begin{center}
\includegraphics[width=1.0\textwidth]{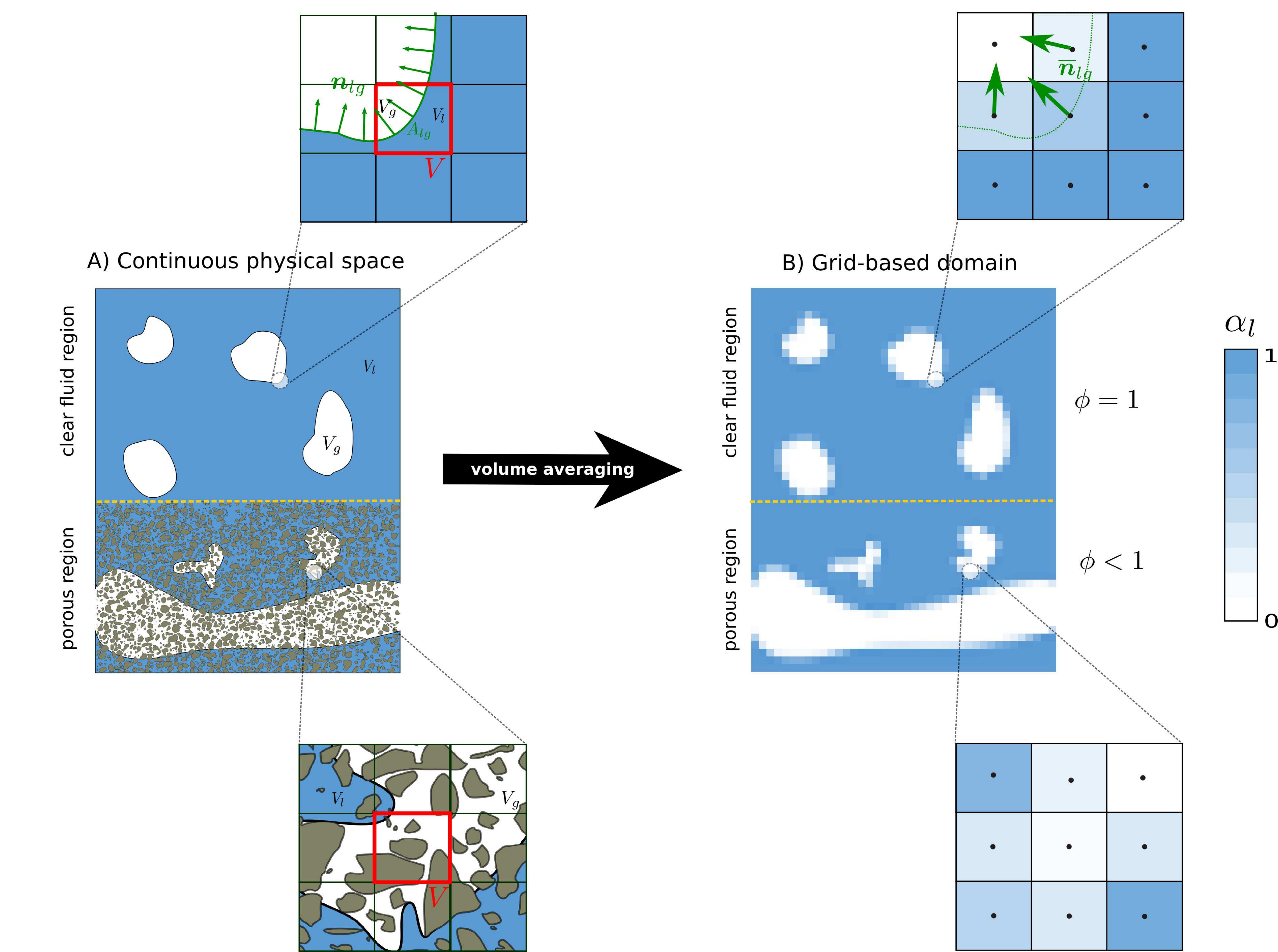}
\caption{Distribution of the fluid phases in (a) the continuous physical domain, (b) the discrete Eulerian grid. \label{fig:FigureVOF}}
\end{center} 
\end{figure}

This section presents the basic hydrodynamic laws that govern multiphase flow at the pore scale. The domain is decomposed into three disjoint subsets: a solid phase $V_s$,  a wetting liquid phase $V_l$, and a non-wetting gas phase $V_g$ which is separated from $V_l$ by the interface $A_{lg}$ (see Figure \ref{fig:FigureVOF}A). Although the fluids are referred to as liquid and gas (or wetting and non-wetting), the derivation and resulting model are valid for any incompressible, immiscible fluid pair including liquid-liquid and liquid-gas systems.

Each fluid phase is assumed to be Newtonian and incompressible. Therefore, mass conservation in each phase dictates
\begin{equation}\label{Eq:phaseMass}
\begin{aligned}
 \nabla \cdot \boldsymbol{v}_i =0 & & \text{in $V_i$,} \quad i=l,g ,
 \end{aligned}
\end{equation}
where $\boldsymbol{v}_i$ is the velocity of phase $i$. Mass conservation at the fluid-fluid interface yields
\begin{equation}\label{Eq:BC12}
\rho_l\left(\boldsymbol{v}_l-\boldsymbol{w}\right)\cdot \boldsymbol{n}_{lg}=\rho_g\left(\boldsymbol{v}_g-\boldsymbol{w}\right)\cdot \boldsymbol{n}_{lg} \hspace{0.5cm} \text{at $A_{ij}$},
\end{equation}
where $\rho_i$ is the density of phase $i$, $\boldsymbol{w}$ is the velocity of the interface, and $\boldsymbol{n}_{lg}$ is the normal vector to the fluid-fluid interface pointing from the wetting to the non-wetting phase. In the absence of phase change, $\boldsymbol{v}_l=\boldsymbol{v}_g=\boldsymbol{w}$ at the fluid/fluid interface.

Momentum conservation in each fluid yields
\begin{equation}\label{eq:momentum_equation}
\begin{aligned}
0 =-\nabla p_i + \rho_i \boldsymbol{g} + \nabla\cdot\mathsf{S}_i  & & \text{in $V_i$,} \quad i=l,g ,
 \end{aligned}
\end{equation}
where $\boldsymbol{g}$ is the gravity vector, $\mathsf{S}_i=\mu_i\left(\nabla \boldsymbol{v}_i + \nabla \boldsymbol{v}_i^T\right)$ is the viscous stress tensor, and $p_i$ and $\mu_i$ are the pressure and viscosity of phase $i$, respectively. In Eq.~\eqref{eq:momentum_equation}, the inertia terms have been neglected and the momentum balance is described using the Stokes equation. This simplification is common in models of subsurface fluid flow, where flow rates are usually very low \citep{Bear1972}. The Stokes equation is adopted for simplicity in the derivation of the micro-continuum momentum equation. For simplicity, inertial effects will be integrated at the end of the derivation based on the full Navier-Stokes equation. 

Finally, momentum conservation at the fluid-fluid interface yields
\begin{equation}\label{Eq:BCMom12}
\left[p_l\mathsf{I}-\mathsf{S}_l\right]\cdot\boldsymbol{n}_{lg}=\left[p_g\mathsf{I}-\mathsf{S}_g\right]\cdot\boldsymbol{n}_{lg}+\sigma\kappa\boldsymbol{n}_{lg} \hspace{0.5cm} \text{at $A_{lg}$},
\end{equation}
where $\mathsf{I}$ is the unity tensor, $\sigma$ is the fluid-fluid interfacial tension, and $\kappa=\nabla\cdot\boldsymbol{n}_{lg}$ is the interface curvature.

\subsection{Volume averaging: derivation of a single-field formulation\label{subsec:derivation_volume_averaging}}
The mathematical model introduced in the former section is defined on a continuous physical domain. Common computational procedures solve this system of equations by discretizing the continuous domain into an ensemble of subset volumes by using the Finite Volume Method (FVM) \citep{Patankar1980}. In the FVM framework, all the physical variables are averaged over each discrete volume. The averaging process and the discretization refinement level dictate that the control volume can contain the following: one fluid, two fluids, one fluid and a solid phase, or two fluids and a solid phase. Features with characteristic length scales below that of the averaging volume (e.g., the geometry of solid-fluid and fluid-fluid interfaces and the forces exerted onto them) must be described using sub-grid scale representations. In this section, we use volume averaging theorems to identify the form of the multiphase micro-continuum equations.

\paragraph{Volume averaging and single-field variables}
In the FVM, the partial differential equations that describe conservation laws, Eqs.~\eqref{Eq:phaseMass} and~\eqref{eq:momentum_equation}, are transformed into discrete algebraic equations by integrating them over each discrete volume $V$. This operation is carried out using the volume averaging operator,
\begin{equation}\label{Eq:volume_averaging_operator}
\overline{\beta_{i}}=\frac{1}{V}\int_{V_i}\beta_i dV,
\end{equation}
where $\beta_{i}$ is a function defined in $V_i$ ($i=l,g$). As in standard volume averaging theory, we also define a phase averaging operator,
\begin{equation}\label{Eq:phase_averaging_operator}
\overline{\beta_{i}}^{i}=\frac{1}{V_i}\int_{V_i}\beta_i dV.
\end{equation} 

The averages defined by Eqs.~\eqref{Eq:volume_averaging_operator} and~\eqref{Eq:phase_averaging_operator} are related through the porosity field $\phi$ and saturation field $\alpha_l$. The porosity field $\phi$ is defined as $(V_l+V_g)/V$, i.e., the volume occupied by both fluids divided by the control volume $V$, such that:
\begin{equation}
\phi=
    \begin{cases}
        1, & \textnormal{in solid-free regions,}\\
        \left]0;1\right[, &  \textnormal{in porous regions.}
    \end{cases}
\end{equation}
The porosity field is the cornerstone of micro-continuum methods because it delineates porous ($0<\phi<1$) and solid-free regions ($\phi=1$). It is intrinsically related to the resolution of the simulation as illustrated in Fig.~\ref{fig:Figure1}. For example, in image-based flow simulations, the control volume size corresponds to the imaging instrument resolution and the porosity field obtained from the gray-scale is used to model sub-voxel micro-structures \citep{Apourvari2014,Soulaine2016,Scheibe2015,Guo2018,Singh2019,AbuAlSaud2020}. By construction of micro-continuum models, all cells must have non-zero porosity \citep{Soulaine2016a}. Hence, a pure solid phase ($\phi=0$) in the micro-continuum framework is described instead as a very low-porosity, very low-permeability domain ($\phi \approx 0$).

The saturation field $\alpha_l$ is defined as $V_l/(V_l+V_g)$, i.e., the volume of liquid divided by the volume occupied by both fluids within the control volume, such that
\begin{equation}\label{Eq:saturation_field}
\alpha_l=
    \begin{cases}
        0, & \textnormal{in regions saturated with gas,}\\
        \left]0;1\right[, &  \textnormal{in unsaturated regions,} \\
        1, &  \textnormal{in regions saturated with liquid.} \\
    \end{cases}
\end{equation}
A saturation field $\alpha_l$ such as that described by Eq.~\eqref{Eq:saturation_field} is used in continuum scale simulations of multiphase flow in porous media (where it represents actual saturation) and in pore scale simulations of multiphase flow in solid-free regions that rely on the VOF representation (where it is used to track the evolution of the immiscible fluid-fluid interface). The relationship $\alpha_l+\alpha_g=1$ is always valid and $\alpha_g$ is deduced from the knowledge of $\alpha_l$. The averaging operators defined by Eqs. ~\eqref{Eq:volume_averaging_operator} and ~\eqref{Eq:phase_averaging_operator} are related by $\overline{\beta_{i}}= \phi \alpha_i\overline{\beta_{i}}^{i}$ ($i=l,g$). 

The two-phase micro-continuum approach relies on single-field variables, i.e., unique fluid pressure and velocity fields that are defined throughout the entire grid regardless of the nature of the phases that occupy the cells. The single-field pressure $\overline{p}$ and velocity $\overline{\boldsymbol{v}}$ are defined as weighted sums of the pressure and Darcy velocity in each fluid phase:
\begin{equation}
\overline{p}=\alpha_l\overline{p}_l^l+\alpha_g\overline{p}_g^g,
\label{eq:single_field_pressure}
\end{equation}
and
\begin{equation}
\overline{\boldsymbol{v}} = \phi\left[\alpha_l\overline{\boldsymbol{v}_l}^l+\alpha_g\overline{\boldsymbol{v}_g}^g\right],
\label{eq:single_field_velocity}
\end{equation}
respectively. We note that the use of porosity-weighted values in Eq.~\eqref{eq:single_field_velocity} yields a single-field velocity equal to the sum of the filtration (Darcy) velocities in each phase, $\overline{\boldsymbol{v}}=\overline{\boldsymbol{v}_l}+\overline{\boldsymbol{v}_g}$.

The governing equations solving for $\overline{p}$ and $\overline{\boldsymbol{v}}$ are obtained using a two-step strategy. First, the volume averaging operator, Eq.~\eqref{Eq:volume_averaging_operator}, is applied to the continuity equations, Eq.~\eqref{Eq:phaseMass}, and to the momentum equations, Eq.~\eqref{eq:momentum_equation}, leading to two pairs of partial differential equations solving for $\alpha_i$, $\overline{\boldsymbol{v}_i}^i$, and $\overline{p}_i^i$ ($i=l,g$).  Second, pairs of phase-averaged equations are added to each other to form the governing equations for the single-field variables. In the averaging process, the volume averaging operator is applied to spatial differential operators (gradient and divergence). This operation is not straightforward because integrals and derivatives can not be interchanged in volumes that contain interfaces including fluid-fluid and fluid-solid interfaces. This is achieved using the spatial volume averaging theorems \citep{Howes1985,Whitaker1999},
\begin{equation}\label{Eq:volume_average_theorem}
    \begin{aligned}
        \overline{\nabla\beta_{i}}&=\nabla\overline{\beta_{i}}+\frac{1}{V}\int_{A_{ij}}\beta_i \boldsymbol{n}_{ij} dA+\frac{1}{V}\int_{A_{is}}\beta_i \boldsymbol{n}_{is} dA, \\
        \overline{\nabla\cdot\boldsymbol{\beta}_{i}}&=\nabla\cdot\overline{\boldsymbol{\beta}_{i}}+\frac{1}{V}\int_{A_{ij}}\boldsymbol{\beta}_i \cdot\boldsymbol{n}_{ij} dA+\frac{1}{V}\int_{A_{is}}\boldsymbol{\beta}_i \cdot\boldsymbol{n}_{is} dA,
    \end{aligned}
\end{equation}
where $A_{ij}$ is the surface area between the two fluids, $A_{is}$ is the surface area between fluid $i$ and the solid phase, $\boldsymbol{n}_{ij}$ is the normal vector at the fluid-fluid interface pointing from $i$ to $j$, and $\boldsymbol{n}_{is}$ is the normal vector at the solid surface pointing from the fluid to the solid. The surface integral terms in these equations transform the boundary conditions at the discontinuity between the fluid phases and at the solid surface into body forces. In others words, the interfacial conditions are included directly in the partial differential equations that describe the conservation laws in the Eulerian grid. 

\paragraph{Mass balance and saturation equations}
The application of the volume averaging theorem, Eq.~\eqref{Eq:volume_average_theorem}, along with the continuity equations, Eq.~\eqref{Eq:phaseMass}, yields \citep{Whitaker1986a}:
\begin{equation}
    \frac{\partial \phi \alpha_i}{\partial t}+\nabla . \overline{\boldsymbol{v}_i} =0,\quad i=g,l.
    \label{eq:averaged_continuity_i}
\end{equation}

The two-phase micro-continuum framework developed in this paper consists of a set of partial differential equations that solve for the single-field variables $\overline{\boldsymbol{v}}$, $\overline{p}$, and $\alpha_l$. Because the volume-averaged continuity equations, Eq.~\eqref{eq:averaged_continuity_i}, involve averaged phase velocities $\overline{\boldsymbol{v}_i}$ they must be transformed into equations in terms of the micro-continuum single-field variables, namely a total fluid conservation equation and a saturation equation.

The total fluid conservation equation is obtained by summing the two continuity equations and assuming that the porous structure is immobile with time, such that:
\begin{equation}\label{Eq:total_mass}
\nabla\cdot\overline{\boldsymbol{v}}=0.
\end{equation}  
Equation ~\eqref{Eq:total_mass} is a divergence-free velocity that is commonly used together with the momentum equation to derive the pressure equation.

The saturation equation is obtained by first introducing the concept of relative velocity:
\begin{equation}\label{Eq:relative_velocity}
\overline{\boldsymbol{v}_r}=\left(\overline{\boldsymbol{v}_l}^l-\overline{\boldsymbol{v}_g}^g\right).
\end{equation}
From the definitions of single-field and relative velocities, we can show that $\overline{\boldsymbol{v}_l}^l=\phi^{-1}\overline{\boldsymbol{v}}+\alpha_g\overline{\boldsymbol{v}_r}$. Because $\overline{\boldsymbol{v}_l}=\phi \alpha_l \overline{\boldsymbol{v}_l}^l$, the saturation equation can be expressed as: 
\begin{equation}\label{Eq:phaseEq}
\frac{\partial \phi \alpha_l}{\partial t} + \nabla\cdot\left(\alpha_l \overline{\boldsymbol{v}}\right) + \nabla\cdot\left(\phi\alpha_l\alpha_g\overline{\boldsymbol{v}_r}\right)=0.
\end{equation}
In equation ~\eqref{Eq:phaseEq}, the wetting phase saturation $\alpha_l$ is advected by the single-field velocity $\overline{\boldsymbol{v}}$. The third term on the left-hand side is an additional convection term involving the relative velocity $\overline{\boldsymbol{v}_r}$. The saturation equation, Eq.~\eqref{Eq:phaseEq}, is exact, i.e., it is derived from elementary physical principles without any assumptions. However, there is no conservation law to solve for $\overline{\boldsymbol{v}_r}$ and this term must be closed. In the forthcoming discussion, we will see that different descriptions of $\overline{\boldsymbol{v}_r}$ are derived for solid-free ($\phi = 1$) and porous regions ($0\leq \phi < 1$). In the first case, the convection term involving the relative velocity serves to compress the fluid-fluid interface and ensures a sharp transition between the immiscible phases. In the second case, $\overline{\boldsymbol{v}_r}$ is closed by matching Eq.~\eqref{Eq:phaseEq} to the standard saturation equation used in multi-phase Darcy flow solvers.

\paragraph{Momentum equation}
A similar procedure is used to form the multiscale momentum equation. First, the volume averaged equations are derived for each fluid. Then, the two resulting equations are combined to form the single-field conservation law.

The application of the volume averaging theorem, Eq.~\eqref{Eq:volume_average_theorem}, to the Stokes momentum conservation equation for fluid $i$, Eq.~\eqref{eq:momentum_equation}, yields \citep{Whitaker1986a,Lasseux1996,Ishii2011}:
\begin{equation}\label{Eq:volume_average_Stokes}
0=-\nabla \left( \phi \alpha_i \overline{p_i}^i \right)+ \phi\alpha_i \rho_i \boldsymbol{g}
+ \nabla\cdot\left(\phi\alpha_i \overline{\mathsf{S}_i}^i \right) + \boldsymbol{D}_{is}+\boldsymbol{D}_{ij},\quad i=g,l,
 \end{equation}
where the two last terms on the right-hand side,
\begin{equation}
    \boldsymbol{D}_{ik}=\frac{1}{V}\int_{A_{ik}} \boldsymbol{n}_{ik} \cdot \left(-p_i \mathsf{I}+ \mathsf{S}_i\right)dA,
\end{equation}
are the drag forces exerted by phase $k$ on phase $i$. In short, $\boldsymbol{D}_{is}$ reflects the friction of fluid $i$ on the solid surface and $\boldsymbol{D}_{ij}$ reflects interfacial shear between the two fluids. These terms accounts for shear that occurs at scales below that of a control volume; therefore, the description of these terms must differ depending on whether the computational cells contain a porous solid structure ($0\leq \phi < 1$) or fluids only ($\phi = 1$). These drag forces will be derived later on in Section \ref{sec:closure}; for the time being, they are kept in their integral forms. 

The sum of the two phase-averaged momentum conservation equations yields:
\begin{equation}\label{Eq:mom}
0=-\nabla \left(\phi \overline{p}\right) + \phi\rho \boldsymbol{g}
+ \nabla\cdot\left(\phi \overline{\mathsf{S}}\right) + \boldsymbol{D}_{ls}+\boldsymbol{D}_{gs}+ \boldsymbol{D}_{lg}+\boldsymbol{D}_{gl},
 \end{equation}
where $\overline{\mathsf{S}}$ is the single-field shear stress $\left[\overline{\mathsf{S}}  = \mu \left(\nabla \overline{\boldsymbol{v}}+\nabla \overline{\boldsymbol{v}}^T\right)\right]$ and $\mu$ is the average fluid viscosity $\left[\mu=\alpha_l\mu_l+\alpha_g\mu_g\right]$. To form the multiscale momentum equation, we express the sum of the average shear stress at the fluid-solid and fluid-fluid interfaces as the sum of two independent terms: a drag force $\mu k^{-1} \overline{\boldsymbol{v}}$ and a surface tension force $\boldsymbol{F}_c$:
\begin{equation}\label{Eq:drag_Terms}
   -\mu k^{-1} \overline{\boldsymbol{v}} + \boldsymbol{F}_c = \phi^{-1}\left[\boldsymbol{D}_{ls}+\boldsymbol{D}_{gs}+ \boldsymbol{D}_{lg}+\boldsymbol{D}_{gl}\right].
 \end{equation}

Eventually, if the porous structure is immobile, the porosity $\phi$ can be removed from the derivatives and the multiscale single-field momentum equation becomes:
\begin{equation}\label{Eq:microcontinuum-momentum}
0=-\nabla  \overline{p} + \rho \boldsymbol{g}
+ \nabla\cdot\overline{\mathsf{S}} -\mu k^{-1} \overline{\boldsymbol{v}} + \boldsymbol{F}_c.
\end{equation}

\paragraph{Summary of the derivation}
The single-field micro-continuum model for incompressible, immiscible two-phase flow in a rigid porous medium, derived above using volume averaging theory, consists of a set of three partial differential equations, namely a total mass balance equation, Eq.~\eqref{Eq:total_mass}, a saturation equation, Eq.~\eqref{Eq:phaseEq}, and a momentum equation, Eq.~\eqref{Eq:microcontinuum-momentum}, that can be solved for the single-field pressure $\overline{p}$, the single-field velocity $\overline{\boldsymbol{v}}$, and the wetting fluid saturation $\alpha_l$:
\begin{align}
&\nabla\cdot\overline{\boldsymbol{v}}=0,
\nonumber \\
&\frac{\partial \phi \alpha_l}{\partial t} + \nabla\cdot\left(\alpha_l \overline{\boldsymbol{v}}\right) + \nabla\cdot\left(\phi\alpha_l\alpha_g\overline{\boldsymbol{v}_r}\right)=0,
\label{eq:two-phase-microcontinuum} \\
&\frac{1}{\phi}\left( \frac{\partial \rho \overline{\boldsymbol{v}}}{\partial t} +\nabla \cdot \left( \frac{\rho}{\phi}\bar{\boldsymbol{v}} \bar{\boldsymbol{v}} \right)\right)=-\nabla  \overline{p} + \rho \boldsymbol{g}
+ \nabla\cdot\overline{\mathsf{S}} -\mu k^{-1} \overline{\boldsymbol{v}} + \boldsymbol{F}_c. \nonumber
\end{align}
In Eq. ~\eqref{eq:two-phase-microcontinuum}, the momentum equation has been modified from Eq. ~\eqref{Eq:microcontinuum-momentum} to include the inertial effects that were neglected above for clarity. The derivation with these inertial effects follows the same averaging procedure as described above, starting from the Navier-Stokes (rather than Stokes) equation \citep{Vafai1981,Hsu1990,Bousquet-Melou2002,Goyeau2003}. The numerical implementation described in Section \ref{sec:numerical_implementation} for solving Eq.~\eqref{eq:two-phase-microcontinuum} accounts for the inertial effects.  

The set of equations presented above is valid throughout the computational domain regardless of the content of a cell. This characteristic is a fundamental aspect of our multiscale solver. It means that the same equations for multiphase flow and transport can be used in both solid-free and porous regions, unlike in the case of multi-physics solvers that involve mortars \citep{Sun2012a,Baber2016}. This feature allows the proposed solver to be applied in media where the pore space is fully resolved and flow is described using the Navier-Stokes equations (pore scale modeling), in media where pores are not resolved and flow is described using Darcy's law (continuum scale modeling), and in intermediate situations that include both fully resolved solid-free regions and porous regions (hybrid scale modelling) as illustrated in Figure \ref{fig:Figure1}. 

A critical feature of the multiscale solver developed in this paper is that it tends asymptotically to the solution of the two-phase Navier-Stokes equations when used as a pore scale model and to the solution of the two-phase Darcy equations when used as a continuum scale model. This is achieved by defining the relative velocity $\overline{\boldsymbol{v}_r}$, the drag force  $\mu k^{-1} \overline{\boldsymbol{v}}$, and the surface tension force $\boldsymbol{F}_c$. These terms are referred to as multiscale parameters because they describe sub-grid scale information such as the location of the fluid-fluid interface and the hydrodynamic impact of the porous micro-structure. They have a different meaning and a different formulation depending on whether the computational grid blocks contain solid material or not.

\subsection{Closure and multi-scale parameters \label{sec:closure}}
 
In the following, we show how the multiscale parameters $\overline{\boldsymbol{v}_r}$, $\mu k^{-1}$, and $\boldsymbol{F}_c$ can be derived by matching Eq. ~\eqref{eq:two-phase-microcontinuum} to its two desired asymptotic models: in the pore scale limit, the algebraic Volume-of-Fluid method; in the continuum scale limit, the multiphase form of Darcy's law.
 
\paragraph{Algebraic Volume-of-Fluid model in the pore scale limit}
In CFD, the Volume of Fluid (VOF) method \citep{Hirt1981} is a standard approach to track the interface movement of two immiscible fluids in a fixed Eulerian grid. This approach is known to approximate the solution of the physical problem, Eqs.~\eqref{Eq:phaseMass}-\eqref{Eq:BCMom12}, using a Finite-Volume grid. In the VOF approach, a phase indicator representing the volume of fluid in each grid block is used to track the distribution of the fluid phases in the computational domain as illustrated in the upper part of Figure \ref{fig:FigureVOF}B. This phase indicator has the same form as the saturation field $\alpha_l$ defined in the two-phase multiscale micro-continuum model. In cells saturated by the wetting phase, $\alpha_l=1$. In cells that contain the non-wetting phase only, $\alpha_l=0$. Finally, $0<\alpha_l<1$ in cells containing the immiscible interface between both fluids. The VOF approach relies on a single-field formulation of the Navier-Stokes equations to compute the two-phase flow. If a cell of the Finite-Volume grid is considered as a control volume, then all the derivation introduced in the previous section can be used to derive the VOF momentum, mass balance, and saturation equations \citep{Maes2019}.

In the standard VOF approaches, the cells do not contain solid ($\phi=1$). The mass balance and saturation equations, Eqs.~\eqref{Eq:total_mass} and ~\eqref{Eq:phaseEq}, remain, therefore, unchanged. The saturation equation with $\phi=1$ is the equation used in algebraic VOF solvers such as \emph{interFoam}, the VOF solver of the open-source CFD code OpenFOAM\textregistered. There, the relative velocity $\overline{\boldsymbol{v}_r}$ is used as a compression term to force the fluid-fluid interface to be as sharp as possible \citep{RuschePhD2002}. This compression velocity acts in the direction normal to the interface. In the VOF framework, the normal to the fluid-fluid interface is computed using the gradient of the saturation. \citet{RuschePhD2002} proposes a relative velocity oriented in the direction normal to interface with a value based on the maximum magnitude of $\overline{\boldsymbol{v}}$:
\begin{equation}
   \overline{\boldsymbol{v}_r}=C_\alpha \max \left(\left| \overline{\boldsymbol{v}} \right|\right) \overline{\boldsymbol{n}_{lg}} ,
\end{equation}
where $C_\alpha$ is a model parameter used to control the compression of the interface and $\overline{\boldsymbol{n}_{lg}}$ is mean normal vector. For low values of $C_\alpha$, the interface diffuses. For higher values, the interface is sharper, but excessive values are known to introduce parasitic velocities and lead to unphysical solutions. In practice, $C_{\alpha}$ is often chosen between 0 and 4. The mean normal vector $\boldsymbol{n}_{lg}$ is computed by using the gradient of the phase indicator function $\alpha_l$. The relation between these two vectors can be obtained by applying Eq.~\eqref{Eq:volume_average_theorem} to the liquid phase indicator function $\mathds{1}_l$ (a function equal to 1 in $V_l$ and 0 elsewhere) in solid-free regions such that \citep{Quintard1994},
\begin{equation}\label{Eq:LemmaBIS}
\nabla\alpha_l=-\frac{1}{V}\int_{A_{lg}}\boldsymbol{n}_{lg}dA.
\end{equation}
Therefore, 
\begin{equation}\label{Eq:meanNormal}
\overline{\boldsymbol{n}_{lg}}=-\frac{\nabla\alpha_l}{\left|\nabla\alpha_l\right|},
\end{equation}
is a unit vector defined at the cell centers that describes the mean normal to the fluid-fluid interface in a control volume.

Another consequence of the absence of solid in the VOF equations is that the forces describing the shear stresses of the fluids onto the solid surface are null, hence $\boldsymbol{D}_{ls}=\boldsymbol{D}_{gs}=0$. Therefore, the Darcy term in the momentum equation vanishes: 
\begin{equation}
\mu k^{-1} \overline{\boldsymbol{v}} = 0.
\end{equation}

The integration of the shear boundary condition at the fluid-fluid interface, Eq.~\eqref{Eq:BCMom12}, yields a relationship between the mutual shear between the two fluids and the surface integral of the surface tension effects: 
\begin{equation}
    \boldsymbol{D}_{lg}+\boldsymbol{D}_{gl} = \phi \boldsymbol{F}_c =\frac{1}{V}\int_{A_{lg}}\boldsymbol{n}_{lg}\cdot \sigma \kappa dA. 
    \label{eq:int_surf_force}
\end{equation}
This equation cannot be used directly, because the terms under the volume integral require the location and curvature of the fluid-fluid interface within a grid block. This information is unknown in a grid-based formulation for which all the physical variables and forces are averaged on control volumes. In the VOF method, the curvature of the interface $\kappa$ is approximated by a mean interface curvature $\overline{\kappa}$. \citet{Brackbill1992} assumes that the mean curvature of the interface can be approximated by calculating the divergence of the mean normal vector,  $\overline{\kappa}=\nabla \cdot \overline{\boldsymbol{n}_{lg}}$. Because $\overline{\kappa}$ and $\sigma$ are constant within a control volume, they can be pulled out of the integral in Eq.~\eqref{eq:int_surf_force} to obtain (after applying Eq.~\eqref{Eq:LemmaBIS}) the so-called Continuum Surface Force (CSF) formulation \citep{Brackbill1992}: 
\begin{equation}
    \boldsymbol{F}_c = \phi^{-1}\sigma \nabla \cdot\left(\frac{\nabla \alpha_l}{\left| \nabla\alpha_l \right|}\right)\nabla \alpha_l.
\end{equation}

\paragraph{Standard two-phase Darcy model in the continuum scale limit}

In this section, we recall the formulation of the standard two-phase Darcy model that is classically used to describe two-phase flow in porous media at the continuum scale \citep{Muskat1949,Miller1998,Pinder2008}. The model can be derived by applying the volume averaging operators on a Representative Elementary Volume of the porous structure \citep{Whitaker1986a,Lasseux1996}, along the same lines of the derivation in Section \ref{subsec:derivation_volume_averaging}. Unlike the present micro-continuum model, the two-phase Darcy model is a two-field model, meaning that instead of one velocity field describing the flow, there are two velocities ($\overline{\boldsymbol{v}_i}$ with  $i=g,l$) with separate pressure fields ($\overline{p_i}$ with  $i=g,l$). 

The incompressible, immiscible two-phase Darcy model consists of a saturation equation for the wetting phase,
\begin{equation}
    \frac{\partial \phi \alpha_l}{\partial t}+\nabla . \overline{\boldsymbol{v}_l} =0,
\end{equation}
a mass balance equation,
\begin{equation}
    \nabla . \overline{\boldsymbol{v}} =0,
\end{equation}
and two momentum balance equations, one for each phase, 
\begin{align}
    \overline{\boldsymbol{v}_i} = \phi \alpha_i \overline{\boldsymbol{v}_i}^i  &=-\frac{k_0k_{r,i}}{\mu_i}\left( \nabla \overline{p_i}^i  - \rho_i \boldsymbol{g}\right),\quad i=g,l, \nonumber
\\
    &= -M_i\left( \nabla \overline{p_i}^i  - \rho_i \boldsymbol{g}\right),\quad i=g,l,
    \label{eq:Darcy-law}
\end{align}
these can also be written as,
\begin{equation}
    0 = - \nabla \overline{p_i}^i  + \rho_i \boldsymbol{g} - M_i^{-1}\overline{\boldsymbol{v}_i} ,\quad i=g,l,
\end{equation}
where $k_0$ is the absolute permeability of the porous structure, $k_{r,l}$ and $k_{r,g}$ are the relative permeabilities with respect to each fluid (classically represented here as functions of water saturation; more complex formulations exist that account for viscous coupling between the two fluids or for the Klinkenberg effect in the gas phase \citep{Standnes2017,Picchi2018,Guo2018}), and $M_i=\frac{k_0k_{r,i}}{\mu_i}$ are the fluid mobilities. These momentum equations arise from further simplification of the volume averaged Stokes equations, Eq.~\eqref{Eq:volume_average_Stokes}, where the drag forces are combined and described as a Darcy term. Moreover, by relying on scale separation arguments, \cite{Whitaker1986} showed that the viscous dissipative term, $\nabla \cdot \left( \alpha_i \overline{\mathsf{S}}_i^i \right)$, is negligible in comparison to the drag forces. This feature is a fundamental aspect of the multiscale micro-continuum framework because it means that even though the viscous dissipative term is retained in the single-field momentum equation, it naturally vanishes when the computational cells contain solid content. This allows the continuity of stresses between porous and solid-free domains \citep{Neale1974}.

Because it involves four equations and five unknown variables, the two-phase Darcy model is complemented by an additional relationship between the two averaged pressure fields that defines the macroscopic capillary pressure $p_c$:
\begin{equation}
    p_c\left(\alpha_l \right)=\left( \overline{p_g}^g - \overline{p_l}^l\right).
    \label{eq:capillary_pressure}
\end{equation}
For simplicity, we follow the classical approximation that $p_c$ depends only on saturation \citep{Leverett1940,Brooks1964,VanGenutchen1980ASoils}. Alternative formulations have been proposed to account for observed disequilibrium and hysteretic effects in the macroscopic capillary pressure \citep{Hassanizadeh2002,Gray2015,Li2018,Miller2019,Starnoni2020}. 

As the two-phase Darcy model explicitly represents the two phase-averaged velocities, it can be used to derive an expression for the relative velocity $\overline{\boldsymbol{v}}_{r}$ in the porous region. Before going through the derivation, we note that the application of the gradient operator to the definition of the single-field pressure $\overline{p}$, Eq.~\eqref{eq:single_field_pressure}, along with the definition of capillary pressure, Eq.~\eqref{eq:capillary_pressure}, results in:
\begin{align}
    \nabla  \overline{p_l}^l &= \nabla  \overline{p} - \nabla \left( \alpha_g p_c \right),
\nonumber
\\
    \nabla  \overline{p_g}^g &= \nabla  \overline{p} + \nabla \left( \alpha_l p_c \right).
    \label{eq:pressure_gradient}
\end{align}
Based on the equations presented above, the multi-phase Darcy model implies the following expression for $\overline{\boldsymbol{v}}_{r}$:
\begin{align}
\overline{\boldsymbol{v}}_{r} &=\left(\overline{\boldsymbol{v}_l}^l-\overline{\boldsymbol{v}_g}^g\right),
\nonumber
\\
&=-\frac{M_l}{\phi \alpha_l}  \left( \nabla\overline{p_l}^l -\rho_l \boldsymbol{g}\right)+\frac{M_g}{\phi \alpha_g} \left(\nabla \overline{p_g}^g -\rho_g \boldsymbol{g}\right),
\nonumber
\\
&=\phi^{-1} \left[-\frac{M_l}{\alpha_l} \nabla\overline{p_l}^l+ \frac{M_g}{ \alpha_g}\nabla\overline{p_g}^g +  \left( \rho_l\frac{M_l}{ \alpha_l}-\rho_g\frac{M_g}{ \alpha_g} \right)\boldsymbol{g}\right],
\nonumber
\\
&= \phi^{-1} \left[-\left(\frac{M_l}{\alpha_l}- \frac{M_g}{\alpha_g}\right)\nabla  \overline{p} +  \left( \rho_l\frac{M_l}{\alpha_l}-\rho_g\frac{M_g}{\alpha_g} \right)\boldsymbol{g} +\frac{M_l}{\alpha_l}\nabla\left(\alpha_g pc \right)+\frac{M_g}{\alpha_g}\nabla\left(\alpha_l pc \right)\right],
\nonumber
\\
&= \phi^{-1} \left[-\left(\frac{M_l}{\alpha_l}- \frac{M_g}{\alpha_g}\right)\nabla  \overline{p} +  \left( \rho_l\frac{M_l}{\alpha_l}-\rho_g\frac{M_g}{\alpha_g} \right)\boldsymbol{g} +\left(M_l\frac{\alpha_g}{\alpha_l} +M_g\frac{\alpha_l}{\alpha_g} \right)\nabla p_c- \left(\frac{M_l}{\alpha_l} - \frac{M_g}{\alpha_g} \right)p_c \nabla \alpha_l\right].
\label{eq:vr_porous}
\end{align}
In \citet{Soulaine2019} only the first term of the right-hand side was considered, such that the model could not account for gravity or capillary effects in the porous domain. The comprehensive formulation presented in Eq.~\eqref{eq:vr_porous} overcomes these limitations.

A two-phase Darcy model for the single-field velocity $\overline{\boldsymbol{v}}$ is then formed to derive the continuum scale formulation of the drag force $\mu k^{-1}\overline{\boldsymbol{v}}$ and capillary force $\boldsymbol{F}_c$. This is achieved by summing both phase velocities, Eq.~\eqref{eq:Darcy-law}, and using the pressure gradient relationship, Eq.~\eqref{eq:pressure_gradient}. We obtain: 
\begin{align}
    \overline{\boldsymbol{v}} &= \overline{\boldsymbol{v}_l}  + \overline{\boldsymbol{v}_g}, 
    \nonumber
    \\
    &=-M_g\nabla \overline{p_g}^g -M_l\nabla \overline{p_l}^l  +\left(  \rho_g M_g+\rho_l M_l\right) \boldsymbol{g},
    \\
    \nonumber
    &= -\left(M_g+M_l\right)\nabla \overline{p} +\left(  \rho_g M_g+\rho_l M_l\right) \boldsymbol{g} + \left[M_l \nabla \left( \alpha_g p_c \right) -  M_g\nabla \left( \alpha_l p_c \right)\right],
\end{align}
The previous equation can be recast into:
\begin{equation}
    0 = -\nabla \overline{p} +\rho^{*} \boldsymbol{g} - M^{-1}\overline{\boldsymbol{v}}+M^{-1}\left[M_l \nabla \left( \alpha_g p_c \right) -  M_g\nabla \left( \alpha_l p_c \right)\right],
    \label{eq:single-field-darcy-momentum}
\end{equation}
where $M=M_l+M_g$ is the total mobility and $\rho^{*}=\left(\rho_lM_l + \rho_gM_g\right)/\left(M_l + M_g\right)$ is a mobility-weighted average fluid density. This single-field two-phase Darcy equation matches the two-phase micro-continuum momentum, Eq.~\eqref{Eq:microcontinuum-momentum}, if the drag coefficient and the capillary force equal
\begin{equation}
    \mu k^{-1} = M^{-1}=k_0^{-1}\left(\frac{\mu_l}{k_{rl}} +\frac{\mu_g}{k_{rg}}\right)^{-1},
    \label{eq:single-field-kr}
\end{equation}
and
\begin{align}
    \boldsymbol{F}_c & = M^{-1}\left[M_l \nabla \left( \alpha_g p_c \right) -  M_g\nabla \left( \alpha_l p_c \right)\right], \nonumber\\
    & = \left[M^{-1}\left( M_l \alpha_g - M_g \alpha_l \right)\left(\frac{\partial p_c}{\partial \alpha_l}\right) - p_c \right]\nabla \alpha_l,
    \label{eq:Fc-1}
\end{align}
respectively. The single-field relative permeability, Eq.~\eqref{eq:single-field-kr}, is a harmonic average of the two-phase mobilities, in agreement with the proposal of \citet{Wang1993} and \citet{Soulaine2019}. 

Finally, we note that in Eq.~\eqref{eq:single-field-darcy-momentum}, the single-field fluid density $\rho^{*}$ in the buoyant term is a weighted average based on the fluid mobilities, or more exactly, the fractional flow functions, $M_iM^{-1}$. This is a classic concept in multiphase flow in porous media. As shown in Appendix B, a strictly equivalent solution can be derived where $\rho^{*}$ is replaced by $\rho$ in Eq.~\eqref{eq:single-field-darcy-momentum} and the capillary force expression is replaced by:
\begin{equation}
    \boldsymbol{F}_c = M^{-1}\left( M_l \alpha_g - M_g \alpha_l \right) [(\rho_l - \rho_g)\boldsymbol{g} + \nabla p_c ] - p_c \nabla \alpha_l
\end{equation}

\paragraph{Condition at the interface between a clear fluid region and a porous domain}

The multiscale parameters are derived above for solid-free and porous regions. In hybrid scale simulations, however, both regions can exist concomitantly in the computational grid (see Fig. \ref{fig:Figure1}B). Here, a condition at the interface between porous and solid-free domains is proposed. 

First, we note that for single-phase flow the DB equation captures the slip length induced by the continuity of stresses between the two regions \citep{Neale1974}. If the porous matrix has sufficiently low permeability, fluid velocities in the porous domain are near zero and a no-slip condition is recovered at the interface between solid-free and porous regions \citep{Angot1999,Khadra2000,Soulaine2016a}. This enables the use of micro-continuum simulations at the pore-scale using a penalized approach, i.e., the solid phase is described as a low-permeability porous medium.  

\begin{figure}
\begin{center}
\includegraphics[width=0.8\textwidth]{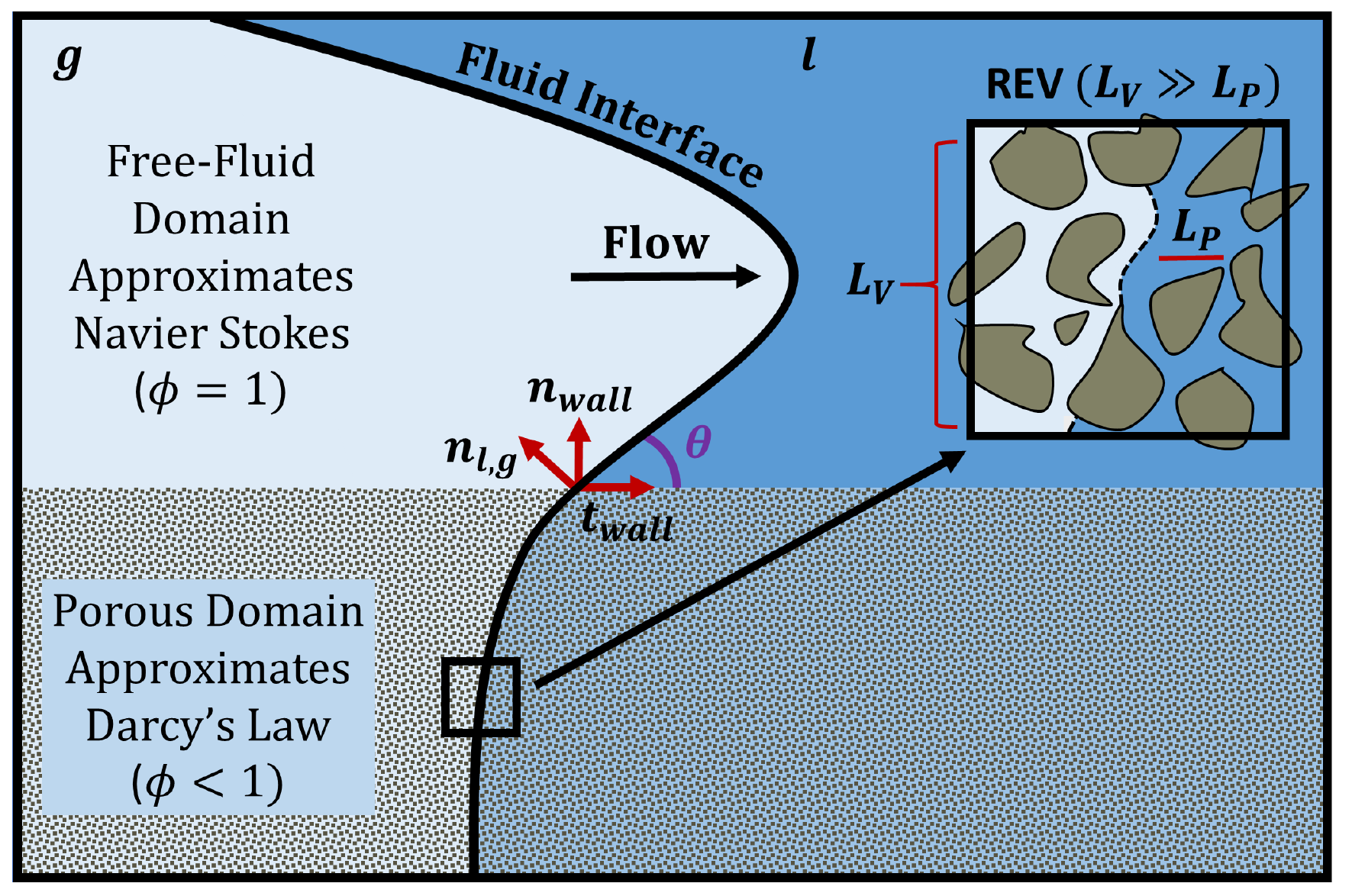}
\caption{Conceptual Representation of the multiphase DB micro-continuum approach. Here $\theta$ represents the contact angle and REV is a Representative Elementary Volume.\label{fig:conceptualDBS}}
\end{center} 
\end{figure}

For two-phase flow, the discontinuity in porosity leads to a change in the form of the surface tension force. Here, we treat this discontinuity by assuming that the fluid-fluid interface of a droplet on a porous substrate forms a contact angle $\theta$ with the solid surface (see Fig.~\ref{fig:conceptualDBS}). The contact angle is an upscaled parameter that depends on various sub-grid scale properties including interfacial energies, surface roughness, and the presence of thin precursor films \citep{Wenzel1936,Cassie1944,Meakin2009}. In the present model, the contact angle is imposed by locally modifying the orientation of the fluid-fluid interface relative to the solid surface \citep{Horgue2014,Soulaine2018,Soulaine2019}. This is achieved by replacing the mean normal vector $\overline{\boldsymbol{n}_{lg}}$ at the interface between the clear fluid and the porous regions by a locally modified normal, $\hat{\boldsymbol{n}}_{lg}$, that satisfies the condition,
\begin{equation}
\hat{\boldsymbol{n}}_{lg}=\cos \theta \boldsymbol{n}_{wall} + \sin \theta \boldsymbol{t}_{wall},
\label{eq:locally_modified_normal}
\end{equation}
where $\boldsymbol{n}_{wall}$ and $\boldsymbol{t}_{wall}$ are the normal and tangent vectors to the porous surface, respectively. The numerical strategy to implement Eq.~\eqref{eq:locally_modified_normal} is described in details in \citet{Horgue2014} and \citet{Soulaine2018}. The effectiveness of this interfacial condition is demonstrated in Section \ref{sec:pore-scale-verification}.


\paragraph{Summary of the multiscale parameters\label{subsec:closure}}

The multiscale parameters $\bar{\boldsymbol{v}}_{r}$, $\mu k^{-1}$, and $\boldsymbol{F}_{c}$ were derived by asymptotic matching to the VOF method in solid-free regions and to the multiphase Darcy model in porous regions. The resulting parameters, therefore, have different forms in different regions. In the porous domains, the multiscale parameters depend on concept such as relative permeability $k_{r,i}$ (also described in terms of fluid mobility, $M_i=k_{r,i}/\mu_i$) and capillary pressure $p_c\left(\alpha_l \right)$.

The relative velocity follows the relation:
\begin{equation}
\bar{\boldsymbol{v}}_{r}=\begin{cases}
C_\alpha \max \left(\left| \overline{\boldsymbol{v}} \right|\right) \frac{\nabla \alpha_l}{\left| \nabla \alpha_l \right|}, & \textnormal{in clear fluid regions,}\\
\phi^{-1} \left[-\left(\frac{M_l}{\alpha_l}- \frac{M_g}{\alpha_g}\right)\nabla  \overline{p} +  \left( \frac{\rho_lM_l}{\alpha_l}-\frac{\rho_gM_g}{\alpha_g} \right)\boldsymbol{g} +\left(\frac{M_l\alpha_g}{\alpha_l} +\frac{M_g\alpha_l}{\alpha_g} \right)\nabla p_c- \left(\frac{M_l}{\alpha_l} - \frac{M_g}{\alpha_g} \right)p_c \nabla \alpha_l\right], &  \textnormal{in porous regions.}
\end{cases}
\end{equation}

The single-field relative permeability is given by:
\begin{equation}
\mu k^{-1}=
    \begin{cases}
        0, & \textnormal{in solid-free regions,}\\
        k_0^{-1}\left(\frac{k_{r,l}}{\mu_{l}}+\frac{k_{r,g}}{\mu_{g}}\right)^{-1}, &  \textnormal{in porous regions.}
    \end{cases}
\end{equation}

The body force $\boldsymbol{F}_{c}$ describes the capillary forces within a computational cell using:
\begin{equation} 
\boldsymbol{F}_{c}=\begin{cases}
-\phi^{-1}\sigma\nabla.\left(\hat{\boldsymbol{n}}_{lg}\right)\nabla \alpha_{l}, & \textnormal{in solid-free regions,}\\
\left[M^{-1}\left( M_l \alpha_g - M_g \alpha_l \right)\left(\frac{\partial p_c}{\partial S}\right) - p_c \right]\nabla \alpha_l, &  \textnormal{in porous regions,}
\end{cases}
\label{eq:surface_tension_forces}
\end{equation}
where the modified normal at the fluid-fluid interface is:
\begin{equation}
\hat{\boldsymbol{n}}_{lg}=\begin{cases}
-\frac{\nabla\alpha_l}{\left|\nabla\alpha_l\right|}, & \textnormal{in solid-free regions,}\\
\cos \theta \boldsymbol{n}_{wall} + \sin \theta \boldsymbol{t}_{wall}, &  \textnormal{at the interface between solid-free and porous regions.}
\end{cases}
\label{eq:modified normal}
\end{equation}

Finally, the single-field fluid density is expressed as:
\begin{equation}
\rho=\begin{cases}
\rho_l \alpha_l + \rho_g \alpha_g, & \textnormal{in solid-free regions,}\\
\left( \rho_gM_g+\rho_lM_l\right)M^{-1}, &  \textnormal{in porous regions.}
\end{cases}
\label{eq:multiscale_density}
\end{equation}

\section{Numerical implementation \label{sec:numerical_implementation}}

The two-phase multi-scale micro-continuum model proposed in this paper is implemented in the open-source CFD platform OpenFOAM\textregistered~ version 7.0 from https://www.openfoam.org. This code is a C++ library
that solves partial differential equations with the finite-volume
method. It handles complex structured and unstructured three dimensional grids by default and
has demonstrated a good scalability for parallel computing of flow
in porous media \citep{Orgogozo2014,Horgue2015,Guibert2015}. One
of its features is that it solves the coupled equations using sequential approaches.
The present section details the solution algorithm developed in this paper. Particularly close attention is paid to the description of the velocity-pressure coupling.

\subsection{Discretization of the equations}

The momentum equation, Eq.~\eqref{eq:two-phase-microcontinuum}, is
transformed into a set of algebraic equations after application of
the finite-volume discretization procedure. The nonlinearity introduced
by the advection term is dealt with by linearizing around the latest
velocity field. The momentum equation is expressed in semi-discrete form (with
successive time levels denoted by $k$ and $k+1$) using a Euler implicit difference scheme:
\begin{equation}
\mathcal{V}\left(\frac{\rho^{k+1}\mathbf{v}_{P}^{k+1}-\rho^{k}\mathbf{v}_{P}^{k}}{\delta t}\right)=-a_{P}^{'}\textbf{v}_{P}^{k+1}+\sum_{NP}a_{NP}^{'}\textbf{v}_{NP}^{k+1}-\nabla p+\rho\boldsymbol{g}+\boldsymbol{F}_{c}-K_{fs}\textbf{v}_{P}^{k+1}.\label{eq:euler_implicit}
\end{equation}
In equation ~\eqref{eq:euler_implicit}, $\mathcal{V}$ and $\delta t$ stand
for the cell volume and time step, respectively. The subscript $P$ indicates
values at the cell center. The coefficients $a_{NP}^{'}$ account
for the influence of neighboring control volumes and primarily include convective and diffusive fluxes across cell faces. $K_{fs}$
corresponds to the exchange of momentum of the fluids with regard to
the solid structure, i.e., the Darcy term in Eq.~\eqref{eq:two-phase-microcontinuum}.
The pressure gradient, buoyancy term, and capillary force are not discretized at this stage.

All explicit source terms other than the pressure gradient, buoyancy term, and capillary force are combined into a single vector, $\textbf{S}=\frac{\mathcal{V}\rho^{k}\mathbf{v}_{P}^{k}}{\delta t}$.
Eq.\ (\ref{eq:euler_implicit}) can then be rearranged as:
\begin{equation}
\left(\frac{\mathcal{V}\rho^{k+1}}{\delta t}+a_{P}^{'}+K_{fs}\right)\textbf{v}_{P}^{k+1}=\sum_{NP}a_{NP}^{'}\textbf{v}_{NP}^{k+1}+\mathbf{S}-\nabla p+\rho\boldsymbol{g}+\boldsymbol{F}_{c}.\label{eq:euler_implicit_bis}
\end{equation}
This equation forms a matrix system that results from the momentum
equation discretization. The term $a_{P}=\left(\frac{\mathcal{V}\rho^{k+1}}{\delta t}+a_{P}^{'}+K_{fs}\right)$
represents the diagonal term of this matrix. Following OpenFOAM\textregistered{}
internal notations \citep{Jasak1996PhD}, the operator $\textbf{H}\left(\textbf{X}\right)=\sum_{NP}a_{NP}^{'}\textbf{X}_{NP}+\textbf{S}$
is introduced and Eq. (\ref{eq:euler_implicit_bis}) becomes:
\begin{equation}
a_{P}\textbf{v}_{P}^{k+1}=\textbf{H}\left(\textbf{v}^{k+1}\right)-\nabla p+\rho\boldsymbol{g}+\boldsymbol{F}_{c}.\label{eq:semi-discretized}
\end{equation}

This semi-discretized form of the momentum balance is used to
form the pressure equation. This is usually achieved by dividing Eq.
(\ref{eq:semi-discretized}) by the diagonal coefficient, $a_{P}$,
and substituting the semi-discretized form of $\textbf{v}^{k+1}$
into the overall mass balance, Eq.~\eqref{Eq:total_mass}, which is a divergence free velocity in the absence of phase change. Finally, the pressure equation can be written as:
\begin{equation}
\nabla.\left(\frac{\textbf{H}\left(\textbf{v}^{k+1}\right)+\rho\boldsymbol{g}+\boldsymbol{F}_{c}}{a_{P}}\right)-\nabla.\left(\frac{1}{a_{P}}\nabla p^{k+1}\right)=0.\label{eq:pressure_equation}
\end{equation}

Further details regarding the discretization procedure in OpenFOAM\textregistered{}
can be found in \citet{Jasak1996PhD} and \citet{Weller1998}. 
The saturation equation, Eq.~\eqref{Eq:phaseEq}, is 
discretized with a Van Leer limiter function for the convection term and a forward Euler scheme for time discretization. 

\subsection{Solution algorithm}

The discretized equations are solved using OpenFOAM\textregistered{}
in a segregated way. In particular, the pressure-velocity coupling
formed by Eqs.~\eqref{eq:semi-discretized} and \eqref{eq:pressure_equation}
is handled by a predictor-corrector algorithm along the same lines
as the Pressure Implicit Splitting-Operator (PISO) algorithm originally
designed by \citet{Issa1985} to solve the transient Navier-Stokes
equations. It is built on the top of the OpenFOAM\textregistered{} VOF solver \emph{interFoam}. The numerical scheme uses the following sequence of steps. First, the saturation equation, Eq.~\eqref{Eq:phaseEq}, is solved explicitly using
the OpenFOAM\textregistered{} implementation of the Flux Corrected
Transport (FCT) theory \citep{Rudman1997} called Multidimensional
Universal Limiter with Explicit Solution (MULES). Details regarding
the MULES algorithm can be found in the  Chapter 5 of \citet{Damian2013}. Second, the boundary values of $\boldsymbol{v}$  and $\boldsymbol{v_r}$ are updated according to Eqs.~\eqref{eq:single_field_velocity} and \eqref{Eq:relative_velocity}. Third, the single-field relative permeability $k^{k+1}$, density $\rho^{k+1}$,
and viscosity $\mu^{k+1}$ are updated using the new value of the saturation
field, $\alpha_l^{k+1}$. The surface tension force, $\boldsymbol{F}_{c}^{k+1}$, is computed
using Eq. (\ref{eq:surface_tension_forces}). Fourth, the velocity field $\mathbf{v}^{*}$ is calculated
by solving implicitly the momentum equation,
\begin{equation}
a_{P}\mathbf{v}_{P}^{*}=\textbf{H}\left(\mathbf{v}^{*}\right)+\rho^{k+1} \boldsymbol{g}+\boldsymbol{F}_{c}^{k+1}-\nabla p^{k},\label{eq:semi_discretised_bis}
\end{equation}
where the gradient of the pressure field is evaluated from the values
computed at the previous time step. This stage is called the momentum
predictor. Fifth, the predicted velocity $\textbf{v}^{*}$ (which does not satisfy the continuity
equation, Eq.~\eqref{Eq:total_mass}) is corrected.
This is achieved by finding $\left(\mathbf{v}^{**},p^{*}\right)$
that obeys,
\begin{equation}
\mathbf{v}_{P}^{**}=\frac{1}{a_{P}}\left[\textbf{H}\left(\mathbf{v}^{*}\right)+\rho^{k+1} \boldsymbol{g}+\boldsymbol{F}_{c}^{k+1}-\nabla p^{*}\right],\label{eq:corrected_velocity}
\end{equation}
\begin{equation}
\nabla.\mathbf{v}^{**}=0.\label{eq:pressure_mass_conservation}
\end{equation}
Based on these two equations, the pressure equation is formulated
as
\begin{equation}
\nabla.\left(\frac{\textbf{H}\left(\textbf{v}^{*}\right)+\rho^{k+1} \boldsymbol{g}+\boldsymbol{F}_{c}^{k+1}}{a_{P}}\right)-\nabla.\left(\frac{1}{a_{P}}\nabla p^{*}\right)=0,\label{eq:pressure_predictor}
\end{equation}
and solved implicitly with a generalized method of Geometric-Algebraic
Multi-Grid (GAMG) embedded in OpenFOAM\textregistered . The corrected
velocity $\textbf{v}^{**}$ is then computed point-wise from Eq. (\ref{eq:corrected_velocity}).
This step (the PISO loop) may be repeated several times
to ensure convergence. \citet{Issa1985} demonstrated that at least
two iterations are required to ensure that the solution of the pressure-velocity
$\left(\textbf{v},p\right)$ coupling satisfies mass conservation.
The resulting values are set to $(\textbf{v}^{k+1},p^{k+1})$ and,
then, the algorithm marches in time as dictated by the imposed Courant-Friedrichs-Lewy (CFL) number.

\subsection{Open-Source Toolbox: \textit{hybridPorousInterFoam}}

The accompanying open-source toolbox follows the implementation described above and consists of four distinct parts: a main directory that includes the licence files, instructional files, release notes, and automated compilation procedures along with three main toolbox sub-directories. The three sub-directories consist of a Solver sub-directory that includes the code for the \textit{hybridPorousInterFoam} solver; a Tutorials sub-directory that includes all the verification and example cases presented in this paper and; and a Libraries sub-directory that includes both the dynamically linked libraries used in the implementation of the penalized contact angle and, also, the \citet{Brooks1964} and \citet{VanGenutchen1980ASoils} porous media models used to calculate the required sub-voxel description of the fluid-fluid interface in terms of relative permeability and capillary pressures (see Appendix A). This last library was obtained from the open-sourced toolbox published in \cite{Horgue2015}. The \textit{hybridPorousInterFoam} toolbox can be accessed from the author's repository (https://github.com/Franjcf). 

\section{Verification \label{sec:verification}}

In this section, the two-phase micro-continuum model is used in various situations for which reference solutions exist. The objective is to verify that the multiscale solver converges effectively towards its two asymptotic limits, namely the two-phase Darcy model at the continuum scale and the VOF formulation at the pore scale.

\begin{table}[!htb]

\begin{minipage}{.5\linewidth}
    \centering
    \setcellgapes{2pt}
    \makegapedcells
    \medskip
\begin{tabular}{||c  c||} 
\hline
 Property & Value \\ [0.5ex] 
 \hline\hline
 Water Density & \SI{1000}{kg.m^{-3}} \\
 \hline
 Water Viscosity &\SI{1e-3}{Pa.s}\\
 \hline
 Air Density & \SI{1}{kg.m^{-3}}    \\
 \hline
 Air Viscosity & \SI{1.76e-5}{Pa.s} \\ 
 \hline
 Oil Density & \SI{800}{kg.m^{-3}}  \\
 \hline
 Oil Viscosity & \SI{0.1}{Pa.s} \\
 \hline
  Gravity & \SI{9.81}{m.s^{-2}} \\
 \hline
\end{tabular}
 \caption{Table of Fluid Properties}
    \label{table:F}
    
\end{minipage}\hfill
\begin{minipage}{.5\linewidth}
 \centering 
 \setcellgapes{2pt}
 \makegapedcells
 \medskip
 \begin{tabular}{||c  c||} 
 \hline
 Model Parameter & Value \\ [0.5ex] 
 \hline\hline
 \(p_0\) & 100 Pa \\
 \hline
 m (Van Genuchten) & 0.5  \\ 
 \hline
 m (Brooks-Corey) & 3  \\ 
 \hline
 \(\beta \) (Brooks-Corey) & 0.5  \\ 
 \hline
\end{tabular}
 \caption{Table of Model Parameters}
    \label{table:M}
\end{minipage}

\end{table}

\subsection{Darcy scale validation \label{sec:Darcy_validation}}

The model’s ability to predict multiphase flow at the Darcy scale is validated against three well-known analytical and semi-analytical solutions. Together, these assessments test for the correct implementation of the relative permeability, gravity, and capillary terms derived in section 2.3. This validation follows the steps outlined in \cite{Horgue2015} for the development and validation of their own multiphase Darcy scale solver: \textit{impesFoam}.  A complete list of parameters used is provided in Tables \ref{table:F} and \ref{table:M}.

\subsubsection{Buckley-Leverett}
\begin{figure}
\begin{centering}
\includegraphics[width=1\columnwidth]{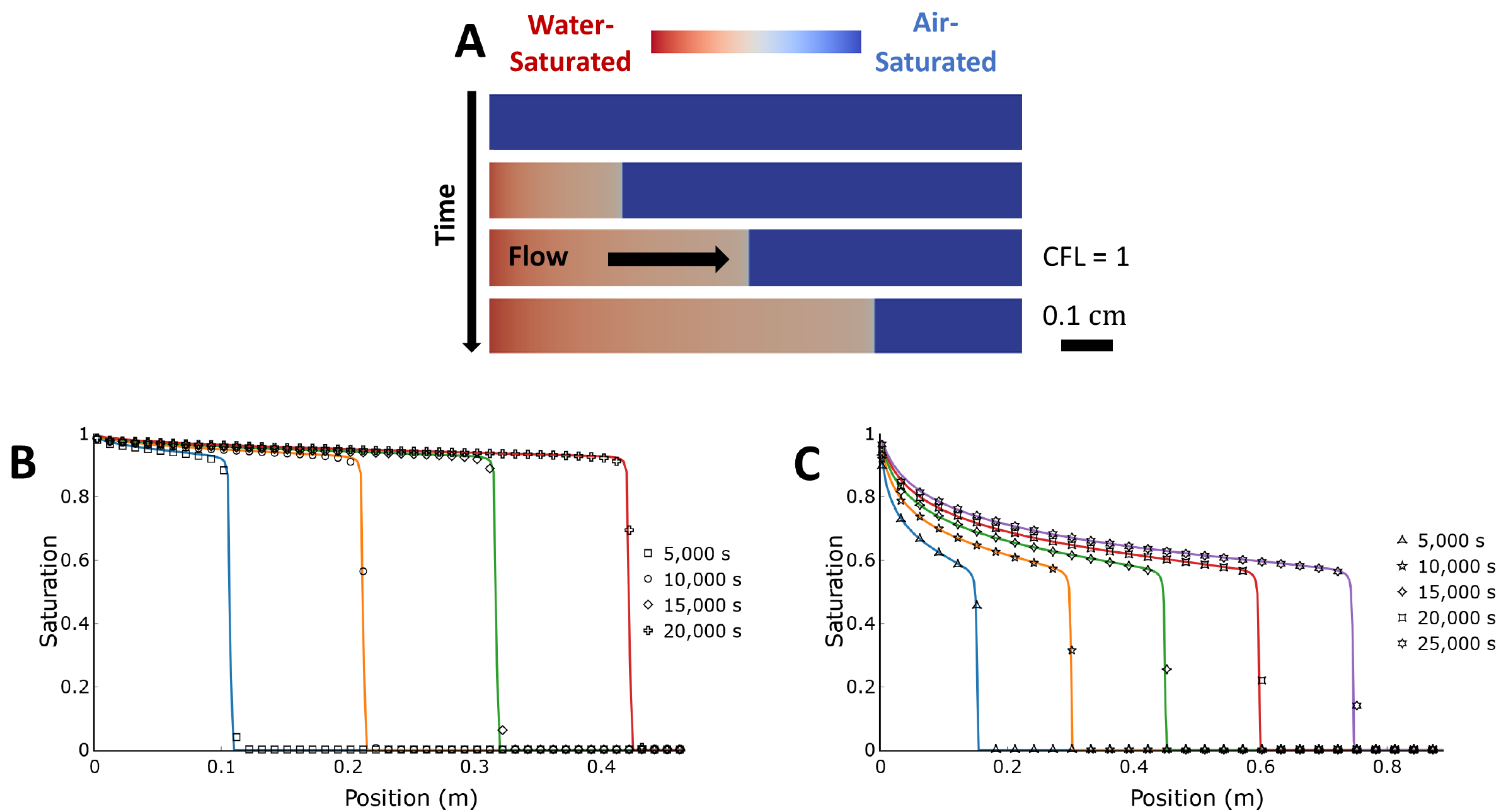}
\par\end{centering}
\caption{Comparison of the time-dependent saturation profiles calculated from our numerical framework and Buckley-Leverett's semi-analytical solution for water injection into air-saturated (B) and oil-saturated (C) reservoirs. Figure A is a visual representation of the water saturation in the reservoir over time. Figures B and C show the semi-analytical (lines) and numerical (symbols) solutions of the system when using the Brooks-Corey and Van Genuchten relative permeability models, respectively.\label{fig:Buckley-Leverett}}
\end{figure}

We first consider the well-established Buckley-Leverett semi-analytical solution for two-phase flow in a horizontal one-dimensional system with no capillary effects (4 m long, 2000 cells, \( \phi = 0.5, k^{-1}_0=1\times10^{11}\) \si{m^{-2}}). In this case, water is injected into an air-saturated reservoir at a constant flow rate with the following boundary conditions: $\textbf{v}_{water}$ = \SI{1e-5}{\meter\per\second}, $\frac{\partial p_{inlet}}{\partial x} =0 \ \si{Pa.m^{-1}}$, and $ p_{outlet} = 0 \ \si{Pa}$. As water flows into the reservoir, it creates a saturation profile that is characterized by a water shock at its front, an effective shock velocity, and a saturation gradient behind said front. Figure \ref{fig:Buckley-Leverett} shows that a good agreement is observed between our numerical solutions and the semi-analytical solutions presented in \cite{Leverett1940} for all three features regardless of the chosen relative permeability model.

\subsubsection{Gravity dominated Buckley-Leverett }

\begin{figure}
\begin{centering}
\includegraphics[width=1\columnwidth]{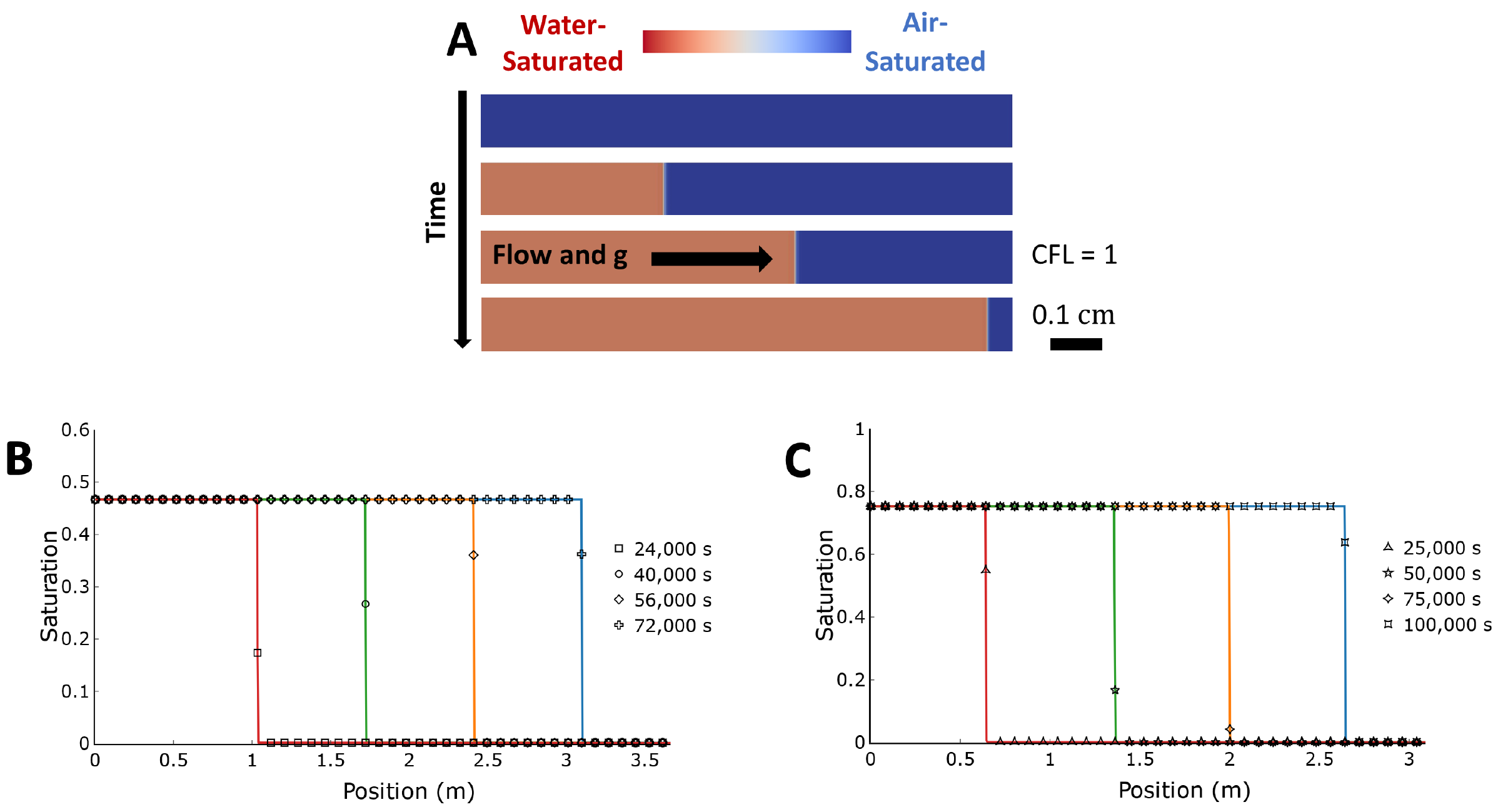}
\par\end{centering}
\caption{Comparison of the time-dependent saturation profiles calculated from our numerical framework and the semi-analytical solution presented in section 4.1.2. Figure A is a visual representation of water saturation in the reservoir over time. Figures B and C show the semi-analytical (lines) numerical (symbols) solutions of the systems parametrized through the Brooks \& Corey and Van-Genuchten relative permeability models, respectively.\label{fig:Buckley-Leverett Gravity}}
\end{figure}

We then tested the exact same air-saturated system, but this time with the addition of gravity in the same direction of the water injection velocity (see Figure \ref{fig:Buckley-Leverett Gravity}). Under these conditions, gravity becomes the dominating driving force and the following equation can be used to calculate the water saturation at the front \citep{Horgue2015}:
\begin{equation}
 \overline{\boldsymbol{v}_l}^l - \frac{k_0 k_{r,l}(\alpha_l^{front})}{\mu_l}\rho_l \boldsymbol{g} = 0,
\label{eq:gravity-front}
\end{equation}
where the symbols are consistent with the ones presented in previous sections. Given the parameters presented in Tables \ref{table:F} and \ref{table:M}, Eq.~\eqref{eq:gravity-front} is solved iteratively to obtain \(\alpha_{l}^{front}  = 0.467  \)  and \(\alpha_{l}^{front}  = 0.753  \) when using the Brooks-Corey and Van Genuchten relative permeability $(k_{r,l})$ models, respectively (Appendix A). Figure \ref{fig:Buckley-Leverett Gravity} shows that our numerical solutions agree with the semi-analytical solutions.  

\subsubsection{Gravity-capillarity equilibrium}

\begin{figure}
\begin{centering}
\includegraphics[width=0.9\columnwidth]{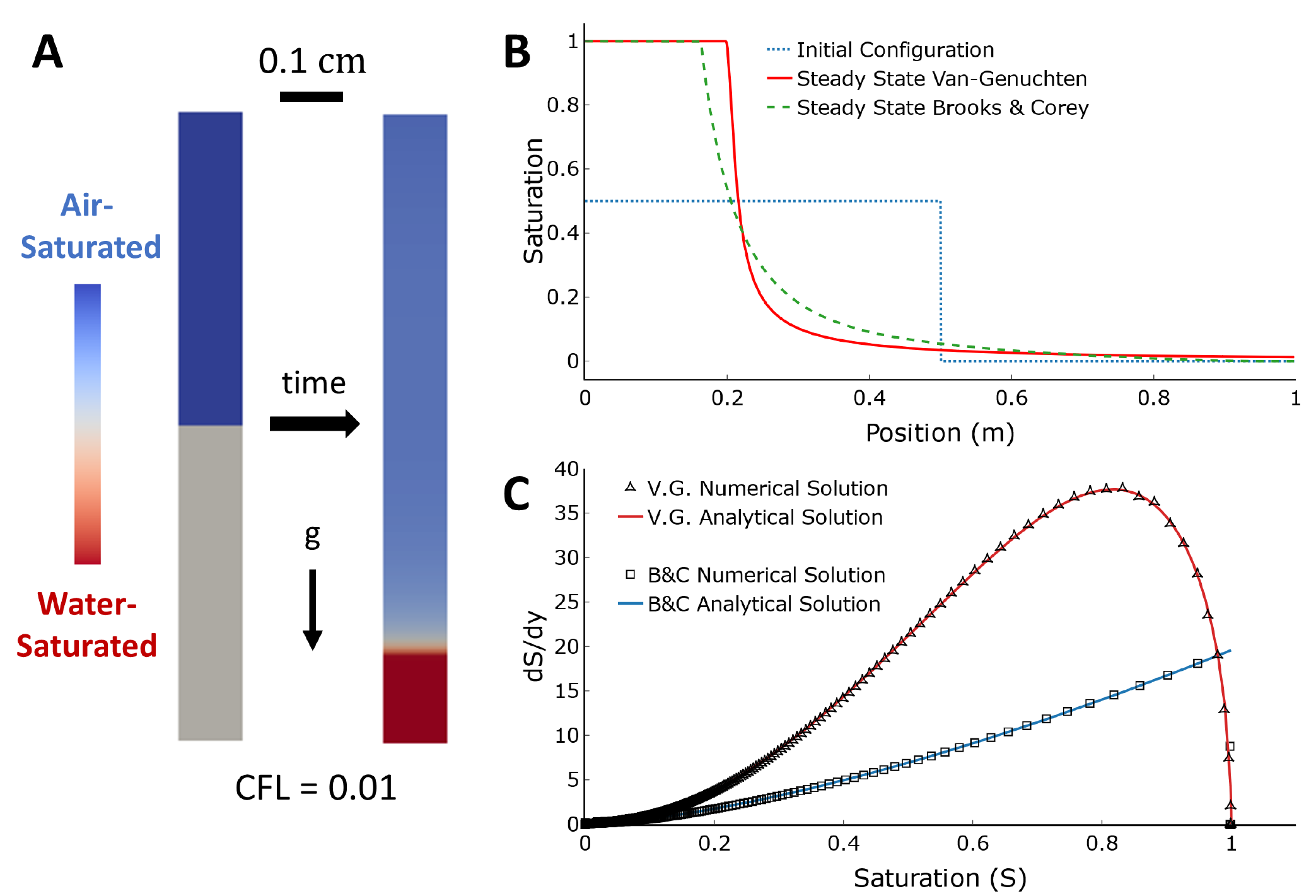}
\par\end{centering}
\caption{Comparison of the steady state water saturation profiles calculated from our numerical framework and the analytical solution shown in equation \ref{eq:gravity-capillarity2}.  Figure A is a visual representation of the initial and final water saturation profiles in the reservoir. Figures B and C show the steady state saturation profiles and the resulting equilibrium saturation gradients for both implemented capillary pressure models, respectively. \label{fig:Gravity-Capillarity Balance}}
\end{figure}

Lastly, we tested the validity of the capillary pressure term derived in Eqs.~\eqref{eq:vr_porous} and ~\eqref{eq:Fc-1} by solving for the steady state saturation profile of a one-dimensional porous column filled with water and air (1 m tall, 1500 cells, \( \phi = 0.5, \ k^{-1}_0=1\times10^{11}\) \si{m^{-2}}). Here, the initial water saturation of the column is set far from its thermodynamic equilibrium in a step-wise fashion: the lower half is partially saturated with water (\(S_{water} = 0.5\)) while the upper half is initially dry as shown in Figure \ref{fig:Gravity-Capillarity Balance}A. To ensure proper equilibriation, both fluids are allowed to flow freely through the column's top boundary, but not through its lower one: $\frac{\partial \textbf{v}_{top}}{\partial y} =0 \ \si{m.s^{-1}.m^{-1}}, \frac{\partial p_{top}}{\partial y} = 0 \ \si{Pa.m^{-1}},
\ \textbf{v}_{bottom} = 0 \ \si{m.s^{-1}}, \ p_{bottom} = 0 \ \si{Pa}$. For this simplified case, the theoretical steady-state can be described as the balance between capillary and gravitational forces, where gravity pulls the heavier fluid (water) downwards while capillarity pulls it upwards. This behaviour can be described by the following equation \citep{Horgue2015}:
\begin{equation}
\frac{\partial p_c}{\partial y} = (\rho_g - \rho_l)g_y,
\label{eq:gravity-capillarity}
\end{equation}
which can be rearranged to yield: 
\begin{equation}
\frac{\partial \alpha_l}{\partial y} = \frac{(\rho_g - \rho_l)g_y}{\frac{\partial p_c}{\partial \alpha_l}}.
\label{eq:gravity-capillarity2}
\end{equation}
This last expression allows for the explicit calculation of the equilibrium water saturation gradient by using the closed-form Brooks-Corey or Van Genuchten capillary pressure models to obtain \(\frac{\partial p_c}{\partial \alpha_l}\) (Appendix A). Figure \ref{fig:Gravity-Capillarity Balance} shows that our numerical model accurately replicates the results obtained from Eq.~\eqref{eq:gravity-capillarity2} regardless of the chosen capillary pressure model.

\subsubsection{Darcy scale application: Oil drainage in a heterogeneous reservoir}

\begin{figure}
\begin{centering}
\includegraphics[width=0.6\columnwidth]{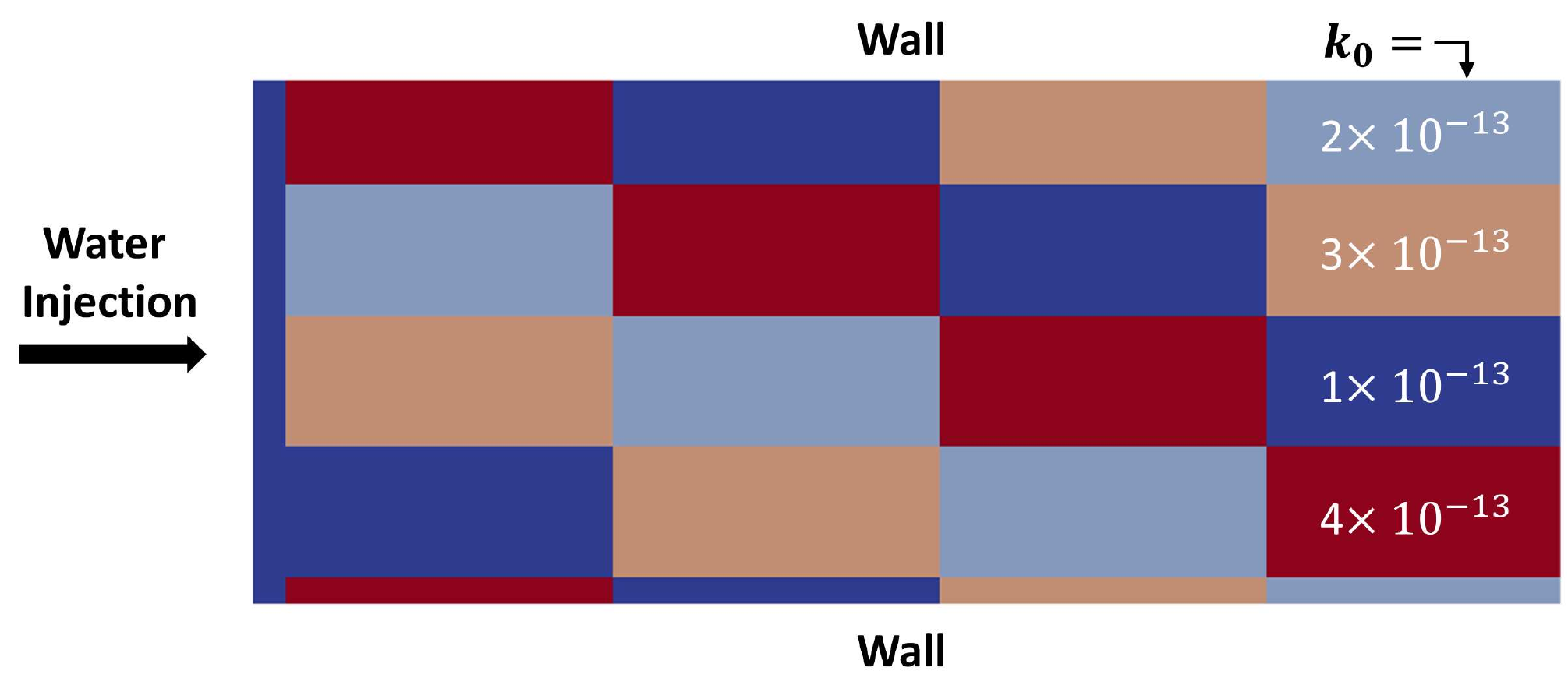}
\par\end{centering}
\caption{Simulation setup for oil drainage within a heterogeneous reservoir. The different colored blocks represent the spatially variable permeability field.  \label{fig:permeabilityField}}
\end{figure}

As an illustration of the applicability of our model to more complex systems at the Darcy scale, we simulate water injection into a heterogeneous oil-saturated porous medium (1 by 0.4 m, 2000 by 800 grid, water injection velocity = \SI{1e-4}{\meter \per \second}, \ $p_{outlet} = 0 \ \si{Pa}$ ). Oil drainage is commonly used in the energy sector, particularly as a form of enhanced oil recovery \citep{Alvarado2010}. Although analytical solutions such as the ones presented above are useful approximations for simple systems, they become greatly inaccurate when modeling complex multi-dimensional systems with spatially heterogeneous permeability. To illustrate this effect, we initialize our reservoir's permeability field as grid of 0.25 by 0.1 m blocks with \(k_0\) values ranging from \SI{1e-13} to \SI{4e-13}{\meter^{2}} (see Figure \ref{fig:permeabilityField}). The relative permeabilities within the reservoir are modeled through the Van Genuchten model with negligible capillary effects  (Table \ref{table:M}). We note that this case was originally presented in \cite{Horgue2015} for the development of \textit{impesFoam}, a solver that uses the Implicit Pressure Explicit Saturation (IMPES) algorithm to solve the two-phase Darcy model, making it a convenient benchmark for comparison with \textit{hybridPorousInterFoam} . 

\begin{figure}
\begin{centering}
\includegraphics[width=1\columnwidth]{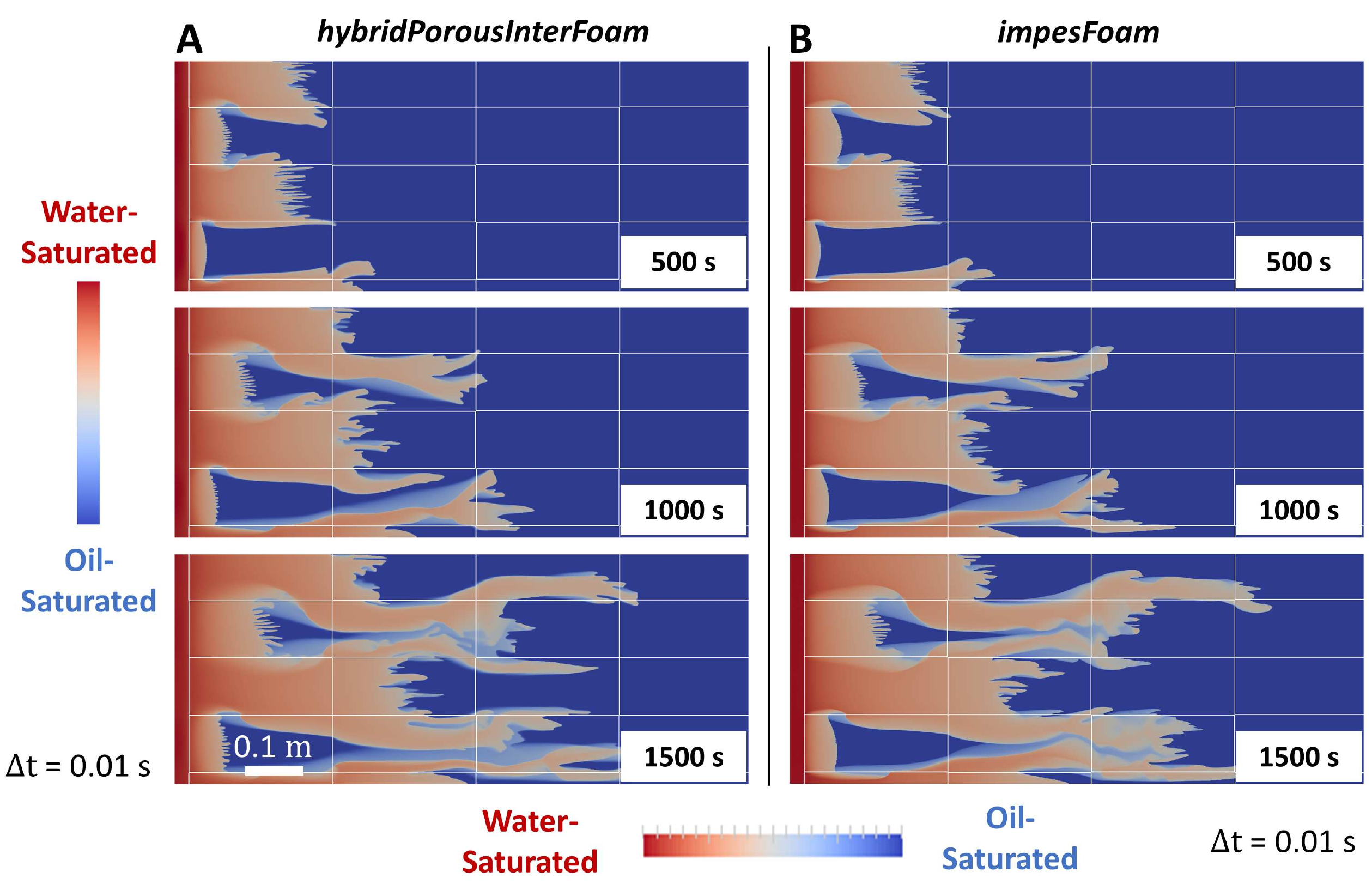}
\par\end{centering}
\caption{Oil drainage in a heterogeneous porous medium solved at the continuum scale using \textit{hybridPorousInterFoam} or \textit{impesFoam}. The white rectangular grid represent the blocks with \(k_0\) values ranging from \SI{1e-13} to \SI{4e-13} {m^{2}} as shown in the previous figure.}   \label{fig:viscousFingering}
\end{figure}

Under the aforementioned conditions, Figure \ref{fig:viscousFingering} shows that the simulations performed with\textit{ hybridPorousInterFoam }and \textit{impesFoam} develop very similar, yet not perfectly equivalent, saturation profiles. Of particular interest is the development of fingering instabilities that form due to the viscosity difference between the two fluids \citep{Url1958,Chen1985}. These instabilities are know to greatly reduce the efficiency of enhanced oil recovery, as they essentially "trap" residual oil behind the main water saturation front (Figure \ref{fig:viscousFingering}). Previous numerical studies have shown that the evolution of viscous fingering is highly dependent on the model's hyper-parameters and/or solver algorithms \citep{Ferrari2013,Riaz2006,Horgue2015,Chen1998,Holzbecher2009}. This characteristic explains why \textit{hybridPorousInterFoam} and \textit{impesFoam} develop slightly different viscous fingering instabilities despite having virtually perfect agreement with the previously-presented analytical solutions: the two solvers rely on entirely distinct sets of governing equations, boundary conditions, discretization schemes, and pressure-solving algorithms (PISO vs IMPES). Nevertheless, this example application shows that our solver can readily simulate complex porous systems that have traditionally been modeled using conventional single-scale Darcy solvers.  

\subsection{Pore scale validation \label{sec:pore-scale-verification}}
Having validated all aspects of the model within the porous domain, we now test our model’s ability to recover known multiphase Navier-Stokes solutions within a non-porous domain. This validation follows the steps used in previous validations of multi-phase CFD solvers by \cite{Horgue2014}, \cite{Xu2017}, and \cite{Maes2019} and involves testing the implementation of the imposed contact angle boundary condition against several well-known numerical and analytical cases. Some of the simulation results obtained with our multi-scale solver are compared with simulations performed using \emph{interFoam}, the algebraic VOF solver of OpenFOAM\textregistered. In the following simulations, we implement a static contact angle as an approximate description of multiphase behaviour at solid interfaces, while noting the existence of more sophisticated formulations including dynamic contact angles with viscous bending or surface roughness \citep{Wenzel1936,Cassie1944,Voinov1976,Cox1986,Whyman2008,Meakin2009}.

\subsubsection{Contact angle on a flat plate}

We first test the implementation of the penalized contact angle within \textit{hybridPorousInterFoam} by initializing several “square” water droplets on a 2-D flat porous plate with negligible permeability (6 by 2.4 mm, 480 by 192 cells, \(k_0^{-1} =\) \SI{1e20}{\meter^{-2}}) and allowing them to reach equilibrium for different prescribed contact angles (\(\theta_{water}\) = \ang{60},\ \ang{90}, \ \ang{150}). These tests are compared against equivalent droplets initialized on conventional non-porous boundaries and solved through \textit{interFoam}. Figure \ref{fig:penalizedContactAngle}A shows excellent agreement between the numerical simulations and the target equilibrium contact angle  \(\theta_{water}\). The lack of a perfectly sharp interface (an intrinsic feature of the VOF method) makes it difficult to accurately measure the contact angle at the solid interface. However, we can confidently state that all our results are within $\ang{5} $ of the target equilibrium contact angle. These tests are virtually identical to the ones shown in \cite{Horgue2014} and are consistent with their results. 

\subsubsection{Capillary rise}

As a second classic test for the correct implementation of multiphase flow at the pore-scale, we model water capillary rise in an air-filled tube (1 by 20 mm, 40 by 400 cells, \(\theta_{water}\) = \ang{45}) and measure the steady-state position of the water meniscus. To ensure a proper numerical setup, the tube's lower boundary is modeled as an infinite water reservoir and its upper boundary as open to the atmosphere. To prevent initialization bias, the meniscus is initialized  about 2 mm lower than the theoretical equilibrium height of 10 mm, which is given by the following equation \citep{Jup1719}:

\begin{equation}
h_{eq.} = \frac{\sigma cos(\theta)}{R \rho_l g_y },
\label{eq:capillaryHeight}
\end{equation}
where $R$ is the tube's radius. We then numerically simulate the system with \textit{hybridPorousInterFoam} and \textit{interFoam}, using impermeable porous boundaries with the former (\(k_0^{-1} =\) \SI{1e20}{\meter^{-2}}) and conventional sharp boundaries with the latter. Figure \ref{fig:penalizedContactAngle}B shows the steady state configurations of both cases, which have a meniscus height of 8.8 mm. According to Eq.~\eqref{eq:capillaryHeight}, this height is equivalent to an imposed contact angle of \ang{52}, a small yet significant difference to the imposed contact angle of \ang{45}. We are not the first to note that \textit{interFoam} (the standard pore scale multiphase flow solver in OpenFOAM\textregistered{} presents minor inaccuracies in its ability to impose a prescribed contact angle \citep{Horgue2014, Grunding2019}. The comparisons presented here show that our solver's accuracy in this regard is similar to that of \textit{interFoam}. 

\subsubsection{Taylor film}

\begin{figure}
\begin{centering}
\includegraphics[width=1\columnwidth]{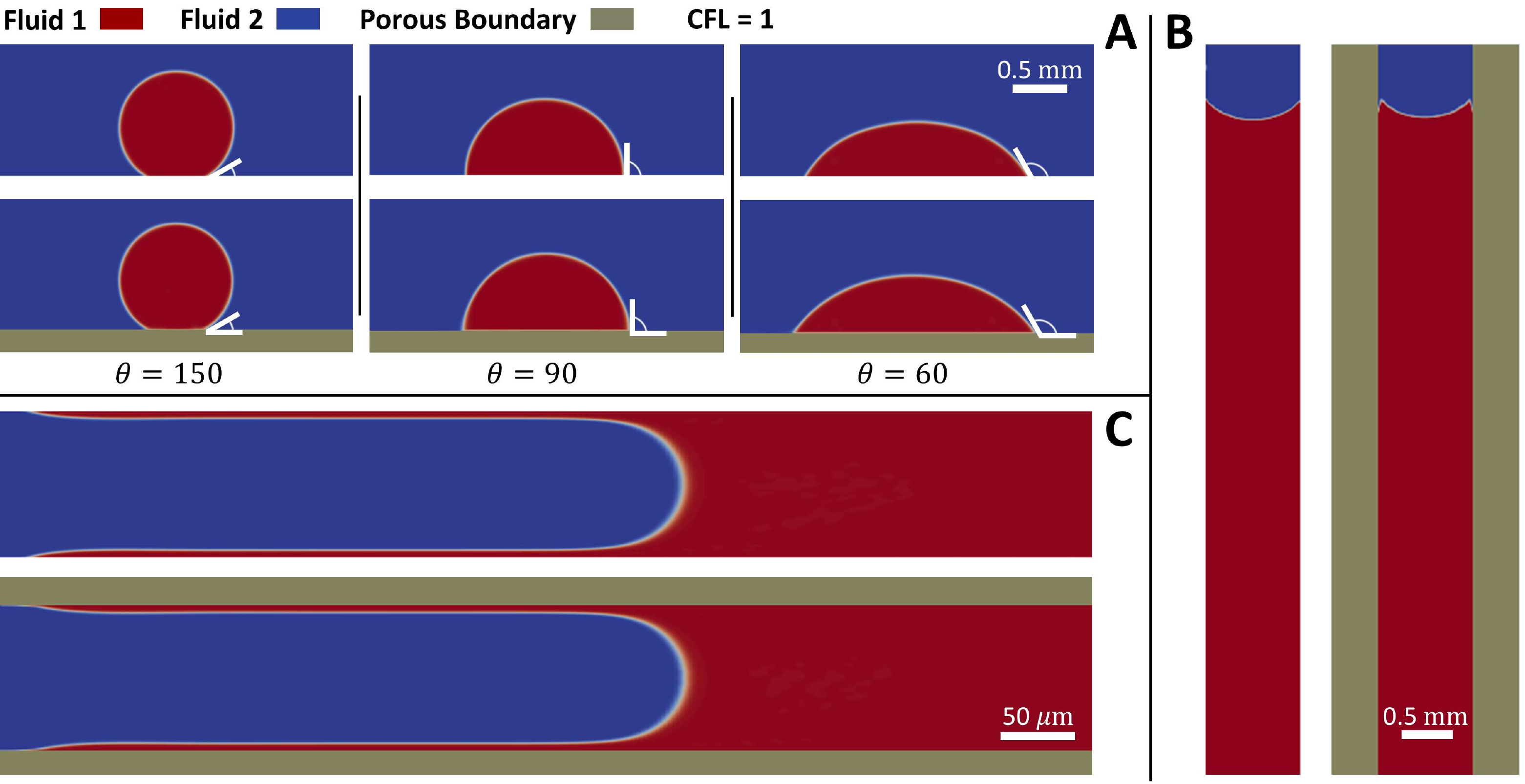}
\par\end{centering}
\caption{Compilation of all test cases performed for the verification of the solver within the Navier-Stokes domain. Parts A, B, and C refer to the experiments described in sections 4.2.1, 4.2.2, and 4.2.3, respectively. When present, the shaded walls show the porous boundaries used in \textit{hybridPorousInterFoam}, as opposed to the standard boundary (no-slip boundary condition at an impermeable wall) using in \textit{interFoam}.For reference and easy comparison, the white lines in Part A show the input equilibrium contact angle.  \label{fig:penalizedContactAngle}}
\end{figure}

We now model the drainage of ethanol (\(\mu_{eth.}\) =\SI{1.2e-3}{Pa.s}, \(\rho_{eth.}\) = \SI{789}{kg.m^{-3}}) by air in a 2-D micro-channel (800 by 100 $\mu$m, 280 by 116 cells, \(\theta_{eth.}\) = \ang{20}, injection velocity ``U" = 0.4 m/s, $p_{outlet}=0 \ \si{Pa}$). Under these circumstances, a  film forms at the channel's boundaries as a result of competing viscous and capillary forces at the solid interface (see Figure \ref{fig:penalizedContactAngle}C). The height of this film is given by the following analytical solution, which we use as a benchmark to verify our numerical simulations \citep{Aussillous2000},
\begin{equation}
\frac{h_{film}}{R} = \frac{1.34 Ca^{2/3}}{1+ 3.35 Ca ^{2/3}},
\label{eq:FilmHeight}
\end{equation}
where $Ca$ is the capillary number defined as $Ca=\frac{\mu_{eth.} U}{\sigma}$. We can solve Eq.~\eqref{eq:FilmHeight} with the given simulation parameters to obtain a film thickness of 4.35 $\mu$m. Simulations of this system performed using \textit{hybridPorousInterFoam} with impermeable porous boundaries ($k_0^{-1} =$ \SI{1e20}{\meter^{-2}}) and \textit{interFoam} with conventional boundaries yield a value of 4.50 $\mu$m, representing a relative error of about 3\% or 0.15 $\mu$m. These tests and their results are consistent with numerical simulations reported by \cite{Graveleau2017} and \cite{Maes2019} using \textit{interFoam}. 

\subsubsection{Pore scale application: Oil drainage in a complex pore network}

\begin{figure}
\begin{centering}
\includegraphics[width=1\columnwidth]{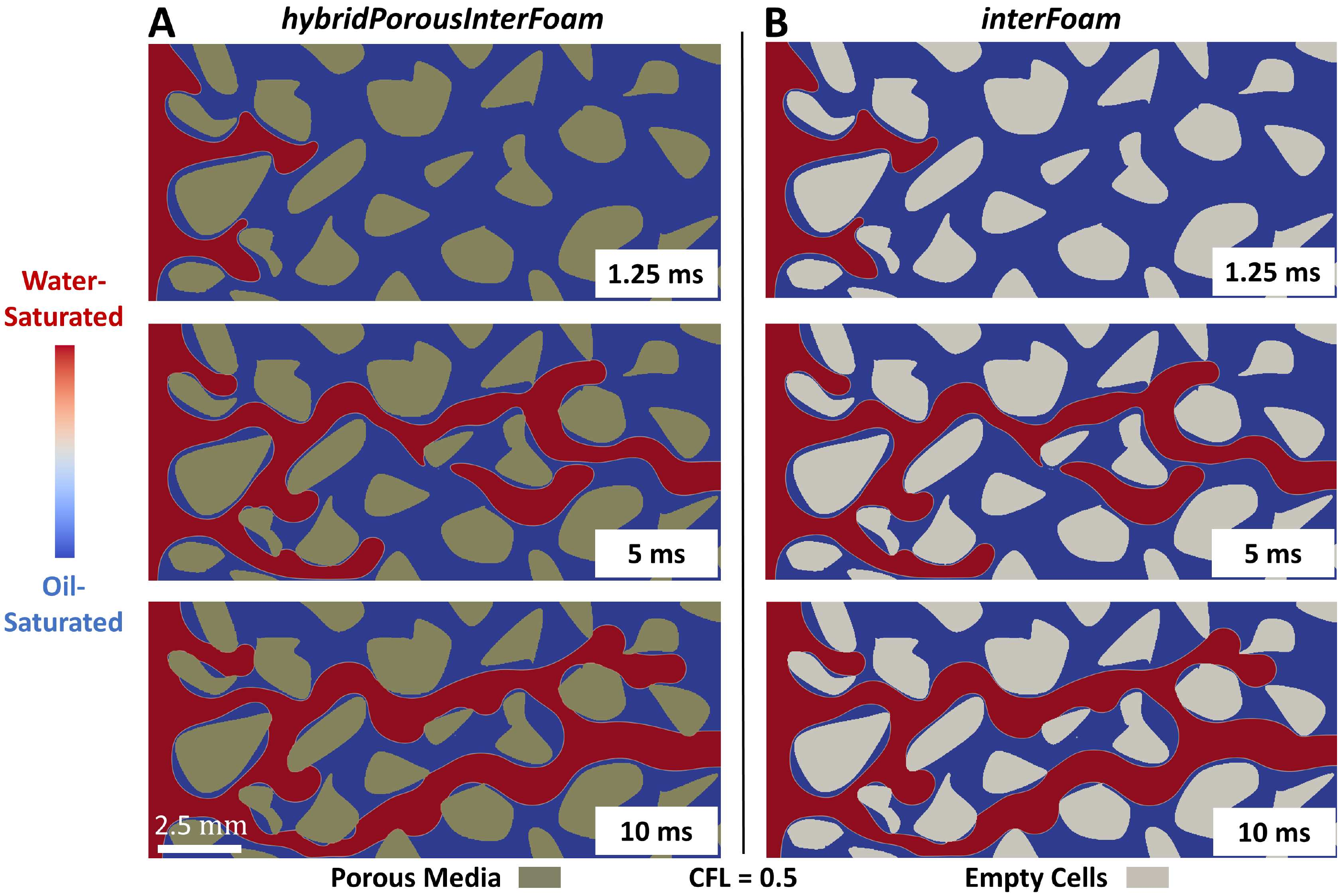}
\par\end{centering}
\caption{Oil drainage in a complex porous medium solved at the pore scale using \textit{hybridPorousInterFoam} and \textit{interFoam}. The shaded sections represent solid grains (modeled using $\phi$ =0.001 and $k_0^{-1} =$ \SI{1e20}{\meter^{-2}} in \textit{hybridPorousInterFoam}) and the blue and red colors represent oil and water, respectively.\label{fig:drainageImbibition}}
\end{figure}

As we did at the end of the Darcy scale verification section, we now illustrate our model's applicability to more complex systems by presenting a simulation of oil drainage, this time at the pore scale. The relevance of the simulated system follows from our previous illustrative problem, as this is simply its un-averaged equivalent at a smaller scale. The complexity of the simulated system (1.7 by 0.76 mm, 1700 by 760 cells, water injection velocity = 0.1 m/s, \(\theta_{oil}\) =\ang{45}, $p_{outlet} = 0 \ \si{Pa}$) stems from the initialization of a heterogeneous porosity field as a representation of a cross-section of an oil-wet rock. Here, the porosity is set to one in the fluid-occupied space and close to zero in the rock-occupied space (See Fig. \ref{fig:drainageImbibition}A). This allows for the solid grains to act as virtually impermeable surfaces ($k_0^{-1} =$ \SI{1e20}{\meter^{-2}}) with wettability boundary condition \citep{Horgue2014}. To verify the accuracy of our solver, we solved an equivalent system with \textit{interFoam} by removing the rock-occupied cells from the mesh and imposing the same contact angle at these new boundaries through conventional methods. 

Figure \ref{fig:drainageImbibition} shows that the results of the two simulations are practically identical, down to the creation of same preferential fluid paths and same droplet snap-off at 5 ms. Nevertheless, there are minor differences in the results, where some interfaces are displaced at slightly different rates than their counterparts (see upper right corner at 10 ms). We attribute these slight differences to the differing implementations of the contact angle at the solid boundaries. We invite the interested reader to find this case in the accompanying toolbox and to refer to the extensive literature on this topic for further discussion on numerical and experimental studies of drainage and imbibition \citep{Lenormand1988NumericalMedia,Ferrari2013,Datta2014,Roman2016,Zacharoudiou2018, Liu2019PreferentialEffects,Singh2019} .

\section{Hybrid Scale Applications \label{sec:applications}}

The complete body of work presented in the previous two sections verifies the capability of our model to perform simulations of multiphase flow in complex porous media at the pore and continuum scales.  We now show how \textit{hybridPorousInterFoam} makes the simulation of hybrid scale multiphase systems a fairly straightforward endeavor, a process that has proven quite challenging to perform with conventional methods. The main challenge when modeling these systems can be summarized by the following question: How can we rigorously model the porous interface between coupled Navier-Stokes and Darcy scale domains? Although this is still an open question, we attempt to approximate an answer by guaranteeing three of its necessary components in the present micro-continuum framework: first, mass conservation across the interface; second, continuity of stresses across the interface and; third, a wettability formulation at the interface. The first two components are intrinsic features of the solver which have been proven necessary and sufficient to accurately model single phase flow in hybrid scale simulations \citep{Neale1974} and have been used as closure conditions in previous multiphase models \citep{Lacis2017,Zampogna2019ModelingMedia}. The latter, as explained in the pore scale validation section, is roughly approximated through a constant contact angle boundary condition. We recognize that these components represent an approximation to the complete description of the boundary. Nevertheless, to the best of our knowledge, there does not exist a better way to model this interfacial behaviour, a testament to the novelty and potential of the proposed modeling framework. 

The following illustrative cases are used to show our model's capability to simulate multiphase systems at the hybrid scale. They are also included as tutorial cases in the accompanying toolbox. 

\subsection{Wave propagation in a coastal barrier}

Coastal barriers are used throughout the world to prevent flooding, regulate water levels, and protect against inclement weather \citep{Morton2002}. Accurate simulations of water interaction with these barriers is challenging as it requires predicting the behavior of open water at large scales (Navier-Stokes) while also resolving small-scale multiphase effects within the barrier itself (Darcy flow). 

\begin{figure}
\begin{centering}
\includegraphics[width=0.65\columnwidth]{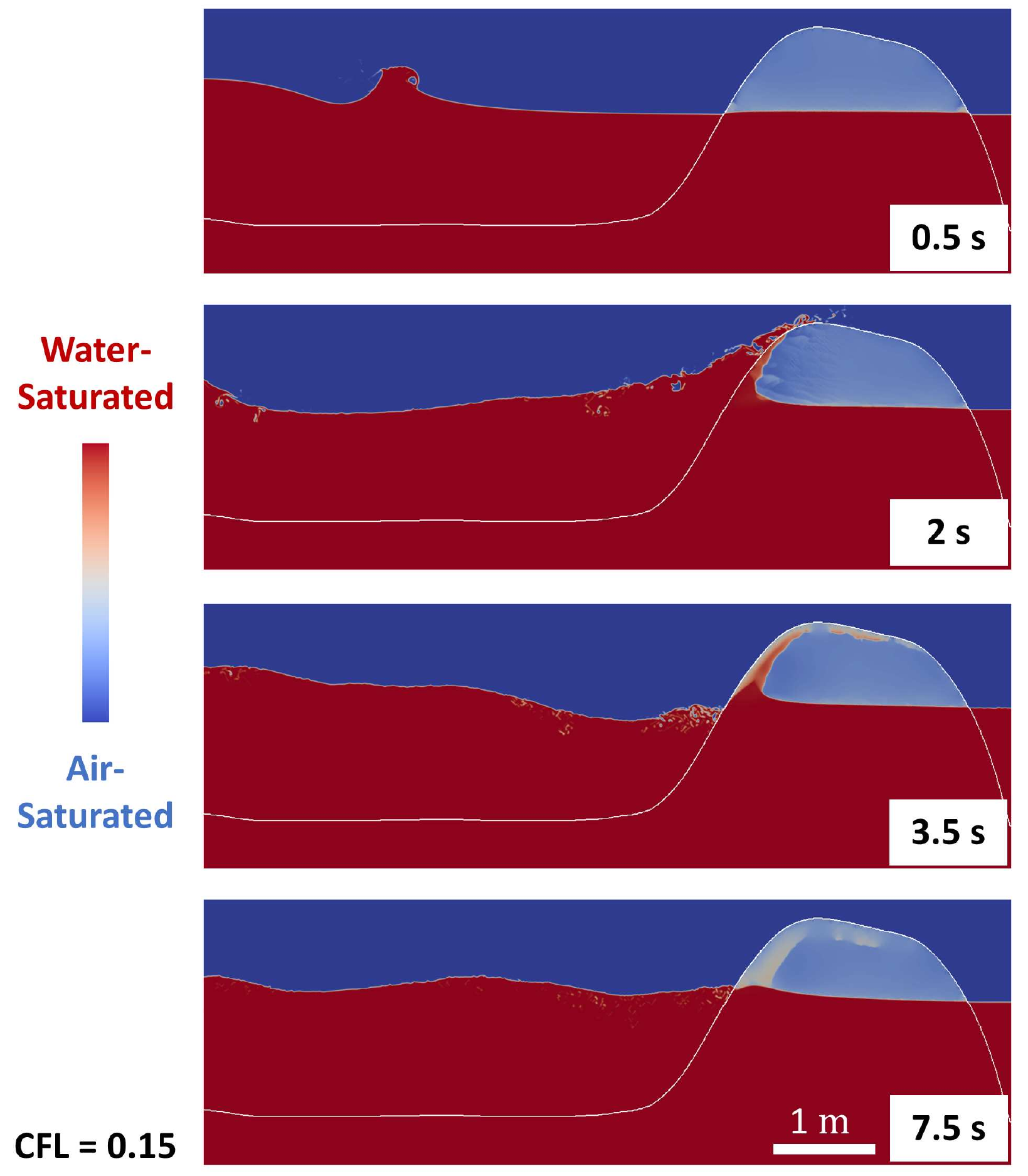}
\par\end{centering}
\caption{Coastal barrier simulation at different times. The thin white line represents the boundary of the porous solid: $\phi$  and $k_0^{-1}$ are set to 0.5 and \SI{5e7}{\meter^{-2}} below said line and to 1 and 0 above it, respectively. \label{fig:coastalBarrier}}
\end{figure}

We created a two-dimensional coastal barrier (8.3 by 2.7 m, 1660 by 540 cells) by initializing a heterogeneous porosity field in the shape of a barrier ($k_0^{-1} =$ \SI{5e7}{\meter^{-2}}, $\phi_{barrier}$ = 0.5) and setting the water level such that it partially covers the barrier (see Figure \ref{fig:coastalBarrier}). In this particular case, we chose not to impose a contact angle at the barrier-water interface as its effects would be minimal when compared to macroscopic gravitational effects (Bond Number $= \frac{\Delta \rho (\textnormal{Length Scale})^2 g_y}{\sigma} >> 1$). To ensure proper initialization, we allowed the water saturation profile on the above-water section of coastal barrier to reach its capillary-induced steady state (similarly to the capillary rise simulation presented in section 4.1.3). This process was modeled using the Van Genuchten relative permeability and capillary pressure models (m = 0.8, $p_0$ = 1000 Pa). After equilibration, we started the simulation by initializing a wave as a square water extrusion above the water surface. To ensure proper wave propagation behavior, we tuned the simulation's numerical parameters (discretization schemes, linear solvers, time stepping strategy) according to the guidelines established in \cite{Larsen2019PerformanceWaves}.

The results from this simulation show that we can simultaneously model coupled wave and Darcy dynamics. The snapshots shown in Figure \ref{fig:coastalBarrier} illustrate how water saturation within the porous domain is controlled by the crashing of waves, gravity, and capillary effects. The associated wave absorption and dissipation cycle brought about by the porous structure is repeated every few seconds with lowering intensity until the initial configuration is eventually recovered. To the best of our knowledge, Figure \ref{fig:coastalBarrier} shows the first existing numerical simulation coupling multiphase flow with real capillary effects at two different scales without the use of different meshes, solvers, or complex interfacial conditions. Other models such as \textit{olaFlow} have been developed to simulate similar wave dynamics with coastal barriers \citep{Higuera2013}. Many of these models rely on the assumption that the pores within the coastal barrier are large ($>$10 cm), meaning that they can reasonably ignore capillary effects within the porous medium. Contrastingly, our model makes no such assumption, meaning it can also be used to model coastal barriers with arbitrarily small pores (such as in sand or gravel structures) and also should be applicable to other types of groundwater-surface water interaction \citep{Maxwell2014}.

\subsection{Drainage and imbibition in a fractured microporous matrix}

A second conceptually similar, yet completely different hybrid scale application of \textit{hybridPorousInterFoam} involves the injection of fluids into fractured porous materials. Accurately capturing the fluid behavior in these systems is especially challenging due to the fact that it requires accounting for multiphase effects simultaneously within the fracture (Navier-Stokes), in the surrounding microporous matrix (Darcy), and at the porous boundary (the contact angle implementation).

Here, we model drainage and imbibition in a water-wet fracture system, where we inject air into a 90\% water-saturated microfracture in the former and we inject water into a 90 \% air-saturated microfracture in the latter (1.2 by 0.5 mm, 1200 by 500 cells, \(\theta_{water}\) = \ang{45}, fluid injection velocity = 0.1 m s$^{-1}$, $p_{outlet} = 0 \ \si{Pa}$). The relative permeabilities and capillary pressures in the heterogeneously-initiated porous domain (\(\phi_{fracture} = 0.5,\)  $k_0^{-1} =$ \SI{4e12}{\meter^{-2}}) are modeled through the Brooks-Corey model with n = 3, $p_0 = 100 \ \si{Pa}$, and $\beta=0.5$. 

\begin{figure}
\begin{centering}
\includegraphics[width=0.85\columnwidth]{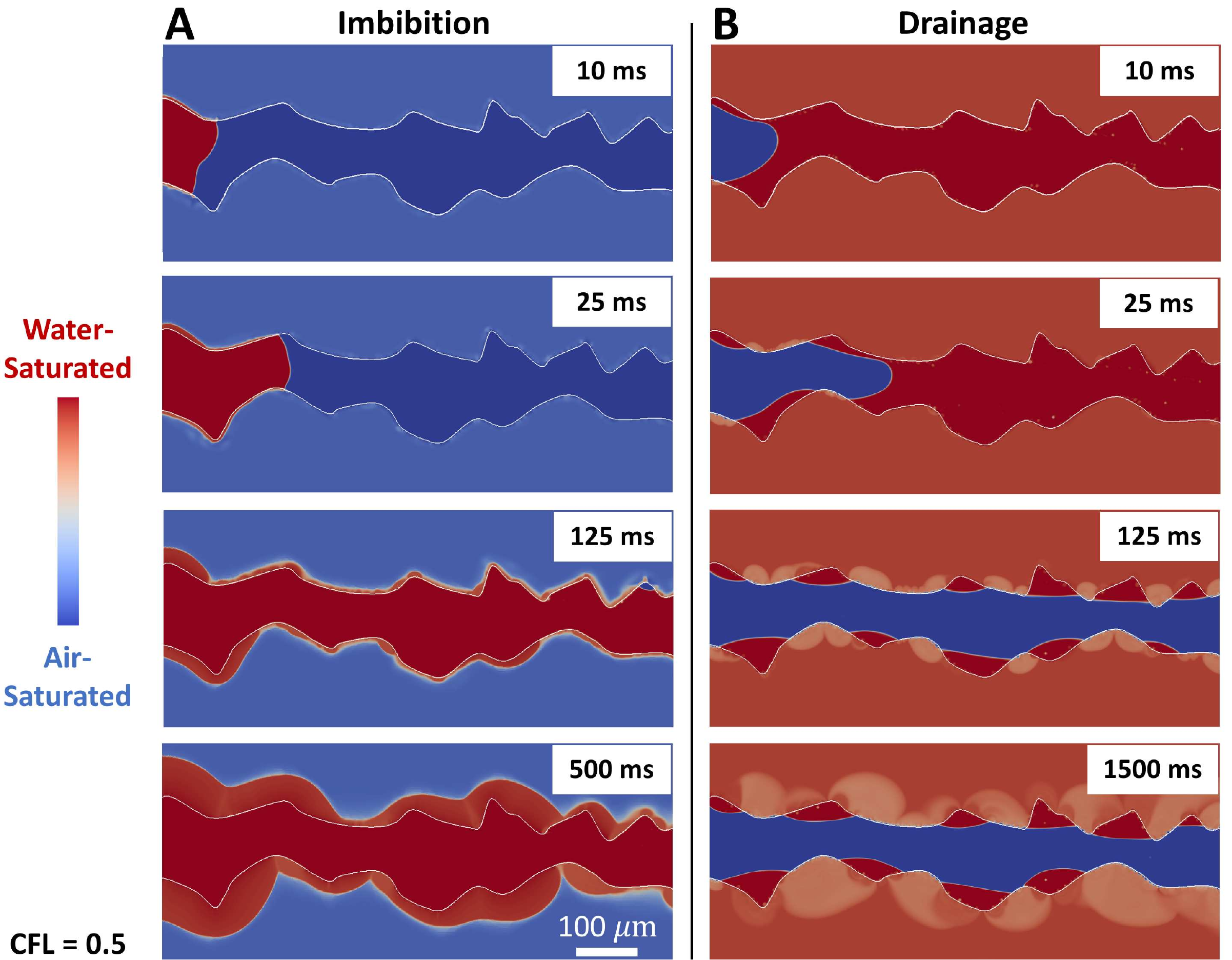}
\par\end{centering}
\caption{Drainage and imbibition in a microporous fracture. Shades of blue and red represent of the degree of air and water saturation, respectively. The thin white line shows the fracture outline (i.e. the fluid-solid interface), which separates the open fracture, (\(\phi =1, k_0^{-1} =0 \)) from the porous fracture walls (\(\phi =0.5, \) $k_0^{-1} =$ \SI{4e12}{\meter^{-2}}) located above and below it.   \label{fig:fracture}}
\end{figure}

Figure \ref{fig:fracture} presents the results of these simulations and illustrates how strongly multi-scale wettability effects can influence simulations results. In both cases, the injected fluid is able to invade the microporous matrix, but the mechanism through which it does is completely different. In the case of water-injection (imbibition), the wetting contact angle boundary condition encourages complete water saturation of the whole fracture such that air is completely displaced by time = 125 ms. Furthermore, throughout the whole process, the microporous capillary pressure acts as an additional driving force for water invasion into the surrounding microporous matrix, leading to the almost complete saturation of the whole system by time = 500 ms.

The drainage case is slightly less intuitive, yet conceptually more interesting. Here, the contact angle and microporous capillary pressures act against the invasion of air into the fracture and into the surrounding porous material, respectively. The result is that the air cannot effectively displace water from the fracture, leading to the trapping of water droplets in fracture ridges. Initially, these droplets act as barriers that prevent air entry into the porous matrix (see time = 125 ms). However, as the flow-induced pressure gradient pushes air into the porous matrix, the water saturation in the pores surrounding the droplets decreases. The system then responds by increasing the capillary pressure at the porous interface, which eventually leads the water droplets to imbibe into the matrix. Lastly, we highlight the clear time scale separation between the imbibition and drainage cases, as the invading interface progresses about three times more slowly within the microporous matrix in the latter case.  

Several similar dual porosity models have been proposed to model the types of effects illustrated in Figure \ref{fig:fracture}, but never in this way or to this degree of detail \citep{Douglas1991AReservoirs,DiDonato2003Streamline-BasedReservoirs}. Many of these models rely on a description of fractures as single-dimensional features with high porosity and permeability values within a pure Darcy scale simulation \citep{Nandlal2019,Yan2016BeyondReservoirs}. Although very useful, many of these simulations ignore the geometric capillary effects and non-linear couplings presented above. Our approach can therefore be seen as the missing link between pore-scale modeling and Discrete Fracture Networks \citep{Karimi-Fard2016} and as a useful tool for the improvement of the transfer function in these large scale models.

\section{Conclusion}

We have successfully derived, implemented, tested, and verified a multiscale model for two-phase flow in porous media. This modeling framework and its open-source implementation \textit{hybridPorousInterFoam} can be used to simultaneously model multiphase flow at two different length scales: a Darcy Scale where sub-voxel fluid-fluid interactions within a porous medium are modeled through relative permeability and capillary pressure constitutive models and a pore scale (or Navier-Stokes scale) where the solid material is non-porous and fluid-fluid interactions are modeled through a continuum representation of the Young-Laplace equation. Furthermore, our model is able to do this through the use of a single momentum conservation equation without the need to define different meshes, separate solvers/domains, or complex interfacial conditions. The proposed framework is an accurate and straightforward way to introduce the physics of two-phase flow in porous media in CFD softwares.

The core derivation of our micro-continuum framework relies solely on fundamental principles and uses the the methods of volume averaging and asymptotic matching to modify and expand the Navier-Stokes equations. Through this study, we showed that our model can successfully simulate multiphase Darcy and Navier-Stokes flow to the same standard as conventional single-scale solvers. The coupling between the two scales at porous interfaces is handled by ensuring mass conservation and continuity of stresses at said boundary, as well as by implementing a constant contact angle wettability condition. We then leveraged all these features to show that our model can be used to model hybrid scale systems such as wave interaction with a porous coastal barrier and drainage and imbibition in a fractured porous matrix. 

Although the proposed formulation represents a significant advance in the simulation of multiscale multiphase systems, we note that further study is required in particular to properly and rigorously model the multi-scale porous interface. The implemented interface, as it stands, has been shown to accurately predict single phase flow into porous media \citep{Neale1974}, impose static contact angles over porous boundaries \citep{Horgue2014}, and approximate multiphase flow in porous media \citep{Lacis2017}. However, its accuracy when modelling multiphase flow at rough porous interfaces is still an open question, as there does not currently exist a rigorous formulation to model such behaviour. The derivation, implementation, and verification of such a boundary condition and the inclusion of erosion, chemical reactions \citep{Soulaine2017,Soulaine2018}, and solid mechanics \citep{Carrillo2019} into this framework will be the focus of subsequent papers. 

\section*{Acknowledgments}

This work was supported by the National Science Foundation, Division of Earth Sciences, Early Career program through Award EAR-1752982. C.S was partially sponsored by BRGM through the TRIPHASIQUE project and by the French National Agency for Research that funded the FraMatI project under contract ANR-19-CE05-0002.

\nomenclature{$\rho_i$}{Density of phase $i$ ($\unit{kg/m^3}$) }%
\nomenclature{$\rho$}{Single-field density ($\unit{kg/m^3}$) }%
\nomenclature{$\boldsymbol{v}_i$}{Velocity of phase $i$ in the continuous physical space ($\unit{m/s}$) }%
\nomenclature{$\overline{\boldsymbol{v}_i}$}{Superficial velocity of phase $i$ in the grid-based domain ($\unit{m/s}$)}%
\nomenclature{$\overline{\boldsymbol{v}_i}^i$}{Phase-averaged velocity of phase $i$ in the grid-based domain ($\unit{m/s}$)}%
\nomenclature{$\overline{\boldsymbol{v}}$}{Single-field velocity in the grid-based domain ($\unit{m/s}$)}%
\nomenclature{$\overline{\boldsymbol{v}_r}$}{Relative velocity in the grid-based domain ($\unit{m/s}$)}%
\nomenclature{$\boldsymbol{w}$}{Velocity of the fluid-fluid interface in the continuous physical space ($\unit{m/s}$)}%
\nomenclature{$V$}{Volume of the averaging-volume ($\unit{m^3}$)}%
\nomenclature{$V_i$}{Volume of phase $i$ in the averaging-volume ($\unit{m^3}$)}%
\nomenclature{$A_{ij}$}{Interfacial area between phase $i$ and $j$ ($\unit{m^2}$)}%
\nomenclature{$\boldsymbol{n}_{lg}$}{Normal vector to the fluid-fluid interface in the continuous physical space}%
\nomenclature{$\overline{\boldsymbol{n}_{lg}}$}{Mean normal to the fluid-fluid interface in the grid-based domain}%

\nomenclature{$p_i$}{Pressure of phase $i$ in the continuous physical space ($\unit{Pa}$) }%
\nomenclature{$\overline{p}_i^i$}{Pressure of phase $i$ in the grid-based domain ($\unit{Pa}$)}%
\nomenclature{$\overline{p}$}{Single-field pressure in the grid-based domain ($\unit{Pa}$)}%

\nomenclature{$p_c$}{Capillary pressure ($\unit{Pa}$) }%

\nomenclature{$\mathsf{S}_i$}{Viscous stress tensor of phase $i$ in the continuous physical space ($\unit{Pa}$)}%
\nomenclature{$\overline{\mathsf{S}_i}^i$}{Viscous stress tensor of phase $i$ in the grid-based domain ($\unit{Pa}$)}%
\nomenclature{$\mathsf{S}$}{Single-field viscous stress tensor  in the grid-based domain ($\unit{Pa}$)}%

\nomenclature{$\sigma$}{Interfacial tension ($\unit{Pa.m}$)}%

\nomenclature{$\kappa$}{Interfacial curvature in the continuous physical space ($\unit{m^{-1}}$)}%

\nomenclature{$\mathsf{I}$}{Unity tensor}%

\nomenclature{$\phi$}{Porosity field}%

\nomenclature{$\alpha_l$}{Saturation of the wetting phase}%

\nomenclature{$\alpha_g$}{Saturation of the non-wetting phase}%

\nomenclature{$\boldsymbol{D}_{ik}$}{Drag force exerted by phase $k$ on phase $i$ ($\unit{Pa/m}$)}%

\nomenclature{$\mu_i$}{Viscosity of phase $i$ ($\unit{Pa.s}$) }%
\nomenclature{$\mu$}{Single-field viscosity ($\unit{Pa.s}$) }%

\nomenclature{$k$}{Apparent permeability ($\unit{m^2}$) }%

\nomenclature{$\boldsymbol{g}$}{Gravity vector ($\unit{m.s^{-2}}$) }%

\nomenclature{$k$}{Apparent permeability ($\unit{m^2}$) }%

\nomenclature{$\boldsymbol{F}_c$}{Surface tension force in the grid-based domain ($\unit{Pa.m^{-1}}) $}%

\nomenclature{$C_{\alpha}$}{Parameter for the compression velocity model}%

\nomenclature{$k_0$}{Absolute permeability ($\unit{m^2}$) }%

\nomenclature{$k_{r,i}$}{Relative permeability with respect to phase $i$ }%

\nomenclature{$M_i$}{Mobility of phase $i$ ($\unit{kg^{-1}m^3s^{-1}}$)}%

\nomenclature{$M$}{Total mobility ($\unit{kg^{-1}m^3s^{-1}}$)}%

\nomenclature{$\theta$}{Contact angle}%

\nomenclature{$\boldsymbol{n}_{wall}$}{Normal vector to the porous surface}%

\nomenclature{$\boldsymbol{t}_{wall}$}{Tangent vector to the porous surface}%

\nomenclature{$p_0$}{Entry Capillary Pressure ($\unit{Pa}$)}%

\nomenclature{$m$}{Van Genuchten Coefficient}%

\nomenclature{$\beta$}{Brooks and Corey Coefficient}%

\printnomenclature

\section*{References}
\bibliographystyle{elsarticle-harv}
\bibliography{mybibliography,referencesFrancisco}






\newpage
\appendix

\section{Relative Permeability and Capillary Pressure Models}

\subsection{Relative Permeability Models}

The two implemented relative permeability models rely on a definition of the effective saturation of the wetting fluid, $\alpha_{l,eff}$, as a function of each fluid's irreducible saturation $\alpha_{i,irr}$:

\begin{equation}
\alpha_{l,eff} =  \frac{\alpha_l-\alpha_{l,irr}}{1-\alpha_{g,irr}-\alpha_{l,irr}}
\end{equation}

The model proposed by \cite{Brooks1964} relates the relative permeability of each phase to the effective saturation through the following relation, where $m$ is a non-dimensional coefficient dictated by porous media properties:

\begin{equation}
k_{rg}= (1-\alpha_{l,eff})^m
\end{equation}
\begin{equation}
k_{rl} = \alpha_{l,eff}^m
\end{equation}
Alternatively, the \cite{VanGenutchen1980ASoils} model relates the relative permeabilities to the wetting fluid's effective saturation in the following way:

\begin{equation}
k_{r,g} = (1-\alpha_{l,eff})^{1/2}(1-\alpha_{l,eff}^{1/m})^{2m}
\end{equation}
\begin{equation}
k_{r,l}= \alpha_{l,eff}^{1/2}(1-(1-\alpha_{l,eff}^{1/m})^m)^{2}
\end{equation}

\subsection{Capillary Pressure Models}

The implemented capillary pressure models rely on slightly different formulations for the effective wetting fluid saturation, $\alpha_{l,pc}$:

\begin{equation}
\alpha_{l,pc} =  \frac{\alpha_l-\alpha_{pc,irr}}{\alpha_{pc,max}-\alpha_{pc,irr}}
\end{equation}
where $\alpha_{pc,max}$ is the maximum saturation of the wetting fluid and $\alpha_{pc,irr}$ is its irreducible saturation. The \cite{Brooks1964} model uses the following expression to calculate capillary pressure within a porous medium:

\begin{equation}
p_c = p_{c,0}\alpha_{l,pc}^{-\beta}
\end{equation}
where $p_{c,0}$ is the entry capillary pressure and $1/\beta$ is a parameter that can be calculated from the pore size distribution. Alternatively, the \cite{VanGenutchen1980ASoils} model uses the relation:

\begin{equation}
p_c = p_{c,0}(\alpha_{l,pc}^{-1/m}-1)^{1-m}
\end{equation}

\section{Alternative formulation of multi-scale parameters}

For consistency with Eq. \ref{Eq:microcontinuum-momentum}, we can recast Eq. \ref{eq:single-field-darcy-momentum} as follows
\begin{equation}
    0 = -\nabla \overline{p} +\rho \boldsymbol{g} - M^{-1}\overline{\boldsymbol{v}}+ (\rho^{*} - \rho) \boldsymbol{g} + M^{-1}\left[M_l \nabla \left( \alpha_g p_c \right) -  M_g\nabla \left( \alpha_l p_c \right)\right],
    \label{eq:single-field-darcy-momentum-2}
\end{equation}

The capillary force $\boldsymbol{F}_c$, then, includes a term in $(\rho^{*} - \rho) \boldsymbol{g}$. After some manipulation, we obtain the following expression (instead of Eq. \ref{eq:Fc-1})
\begin{align}
    \boldsymbol{F}_c & = (\rho^{*} - \rho) \boldsymbol{g} + M^{-1}\left[M_l \nabla \left( \alpha_g p_c \right) -  M_g\nabla \left( \alpha_l p_c \right)\right] \nonumber\\
    & = M^{-1}\left( M_l \alpha_g - M_g \alpha_l \right) (\rho_l - \rho_g)\boldsymbol{g} + M^{-1}\left( M_l \alpha_g - M_g \alpha_l \right) \nabla p_c - p_c \nabla \alpha_l \nonumber\\
    & = M^{-1}\left( M_l \alpha_g - M_g \alpha_l \right) [(\rho_l - \rho_g)\boldsymbol{g} + \nabla p_c ] - p_c \nabla \alpha_l
\end{align}

At equilibrium $(\overline{\boldsymbol{v}_l}  = \overline{\boldsymbol{v}_g} = 0)$, the multiphase Darcy's law yields $(\rho_l - \rho_g)\boldsymbol{g} + \nabla p_c = 0$. Therefore, the equation presented above is consistent with the expectation that at equilibrium the capillary term should be independent of the fluid mobilities and the overall momentum equation should reduce to $0 = -\nabla \overline{p} +\rho \boldsymbol{g} - p_c \nabla \alpha_l$. The preceding derivation suggests an alternative formulation where density is defined identically in the clear fluid and porous regions ($\rho=\rho_l \alpha_l + \rho_g \alpha_g$) while the expression for $\boldsymbol{F}_c$ becomes
\begin{equation}
\boldsymbol{F}_{c}=\begin{cases}
\sigma\nabla.\left(\hat{\boldsymbol{n}}_{lg}\right)\nabla \alpha_{l}, & \textnormal{in the clear fluid regions,}\\
M^{-1}\left( M_l \alpha_g - M_g \alpha_l \right) [(\rho_l - \rho_g)\boldsymbol{g} + \nabla p_c ] - p_c \nabla \alpha_l, &  \textnormal{in the porous regions.}
\end{cases}
\end{equation}
\end{document}